\documentclass[nofootinbib,reprint,amsmath,amssymb,aps]{revtex4-1}
\usepackage[utf8,latin1]{inputenc}
\usepackage{graphicx}
\usepackage{dcolumn}

\usepackage{mathabx}
\usepackage{mathrsfs}  
\usepackage{cases}
\usepackage{bm}
\newcommand{\qm}[1]{``#1''}
\usepackage{hyperref}
\usepackage{stackengine,scalerel}
\hypersetup{colorlinks, linkcolor={red},citecolor={blue},urlcolor={blue}}

\newcommand\overstarbf[1]{\ThisStyle{\ensurestackMath{%
  \stackengine{0pt}{\SavedStyle\mathbf{#1}}{\smash{\SavedStyle*}}{O}{c}{F}{T}{S}}}}

\begin{document}

\title[First-post-Newtonian generation of gravitational waves in Einstein-Cartan theory]{First-post-Newtonian generation of gravitational waves in Einstein-Cartan theory}

\author{Emmanuele Battista$^{1}$\vspace{0.5cm}}\email{emmanuele.battista@kit.edu} \email{emmanuelebattista@gmail.com}
\author{Vittorio De Falco$^{2,3}$}
\email{vittorio.defalco@physics.cz}

\affiliation{$1$ Institute for Theoretical Physics, Karlsruhe Institute of Technology (KIT), 76128 Karlsruhe, Germany\\
$^2$ Department of Mathematics and Applications \qm{R. Caccioppoli}, University of Naples Federico II, Via Cintia, 80126 Naples, Italy\\
$^3$ Istituto Nazionale di Fisica Nucleare, Sezione di Napoli, Complesso Universitario di Monte S. Angelo, Via Cintia Edificio 6, 80126 Napoli, Italy
}

\date{\today}

\begin{abstract}
In this paper we investigate the gravitational-wave generation problem at the first post-Newtonian order in the context of Einstein-Cartan theory by exploiting the Blanchet-Damour formalism.  The quantum intrinsic spin carried by slowly moving, weakly stressed, weakly self-gravitating sources is described geometrically by means of the torsion tensor. We obtain the expression of the source multipole moments with the required accuracy. The analysis of the physical meaning of the lowest-order non-radiative moments and of the asymptotic gravitational waveform is also performed. Eventually, we draw our conclusions and estimate the order of magnitude of the spin contributions in the gravitational-wave signal.
\end{abstract}

\maketitle

\section{Introduction}
\label{sec:intro}
Gravitational-wave (GW) astronomy is an emerging branch of physics which is making  many significant breakthroughs.  The born of GWs observational campaign started  in 2016, when the Laser Interferometer Gravitational-Wave Observatory (LIGO) and the Virgo Collaboration teams announced the first \emph{direct} detection of GWs resulting from the merging of two black holes (BHs) \cite{Abbott2016}. This event, called GW150914, has opened a new era both in astrophysics and in cosmology. In 2017, the first case of GWs originated by colliding neutron stars,   called GW170817, was pinpointed by  LIGO and Virgo  and the consequent emission of  short-gamma-ray bursts  was observed  by INTEGRAL and Fermi Gamma-Ray Burst Monitor \cite{Abbott2017-NS}. This cosmic phenomenon has been a significant milestone in multi-messenger astronomy. A total of fifty GWs events has been spotted to date by  LIGO  and Virgo, as  reported in the Gravitational-Wave-Transient Catalog 2 \cite{Abbott2020}\footnote{For further details see also\\ \href{url}{https://www.ligo.org/science/Publication-O3aCatalog/}.}. This number is destined to  increase,  due to the support of other operating detectors (such as the high-frequency ground-based second-generation interferometer Kamioka Gravitational Wave Detector (KAGRA) \cite{Akutsu2018}), and  near-future \emph{apparati} based both 
on Earth (e.g., the third LIGO detector LIGO-India \cite{Saleem2021} and the third-generation interferometers Einstein Telescope \cite{Punturo2010} and  Cosmic Explorer \cite{Abbott:2016mbw}) and in space (e.g., the low-frequency Laser Interferometer Space Antenna (LISA) \cite{Audley2017}). These devices are expected to  enable  very precise tests of general relativity (GR) in the strong-field regime with an unprecedented sensitivity, which  will range from some tens of Hz to about one kHz for ground-based interferometers and from some $\mu$Hz to about one tenth of Hz for space-based ones \cite{Bailes2021,Miller2019a,AmaroSeoane2012}. Furthermore, we also mention the low-frequency galactic-scale GWs detectors called Pulsar Timing Arrays (PTAs), whose frequency band goes from  $100$ nHz to $1$ nHz \cite{Hobbs2010,Tiburzi2018}. These new technologies  are expected to establish whether GR is the only fundamental theory suited to the description of gravitational interactions or instead extended theories of gravity need to be  introduced. 

In the literature, a plethora of methodologies  aimed at inspecting GWs phenomena  has been conceived.Nowadays, a crucial role in the study of the dynamics of compact binaries is fulfilled by numerical relativity. In 2005, the first successful numerical simulation of the GWs coming from an inspiraling pair of BHs through their merger and final ringdown was achieved \cite{Pretorius2005,Sperhake2014}.
Since then, several international scientific communities developed advanced numerical codes featured by steadily improved and largely-used high performance computing facilities (see review articles \cite{Cardoso2014,Grandclement2007,Faber2012}, for more details). Numerical techniques provide  gravitational waveform templates capable of  both validating the theoretical predictions and fitting the observational data \cite{Berti2016a,Eisenstein2018,Bishop2016,Centrella2007,Barack2018a}. In the current GWs data analysis, two families of inspiral-merger-ringdown approaches are generally employed (see Refs. \cite{Abbott2020a,Abbott2020b}  for more details, and   Ref. \cite{Barack2018a} for a review): (1) \emph{effective-one-body (EOB) models} are  built on post-Newtonian (PN) and perturbation theory results \cite{Buonanno1999,Buonanno2000,Damour2011}, and describe both the dynamics and the waveforms in the time domain of generic  binary systems; (2) \emph{phenomenological patterns (Phenom)} provide a closed-form characterisation of the  frequency-domain waveform  from generic configurations of compact binaries \cite{Ajith2008,Ajith2011,Santamaria2010,Sascha2016a,Khan2016a,Schmidt2012,Schmidt2015}. Moreover, a framework which borrows ideas and techniques from quantum field theory and is extensively used to study the problems of motion and radiation of compact binaries is given by the effective-field-theory approach \cite{Donoghue1994,Burgess2003,Goldberger2006,Foffa2014r}. This method  brings out efficient tools
to compute observable quantities and hence  it is essential in the construction of the template waveforms necessary for the GWs detection \cite{Foffa2017,Foffa2019,Levi2020}.
Furthermore, two different GWs generation frameworks have been developed up to a high PN accuracy: the Will-Wiseman-Pati method, which 
uses a direct integration of the relaxed Einstein equations (DIRE) \cite{Pati2000,Pati2002}, and the Blanchet-Damour formalism \cite{Blanchet-Damour1986,Blanchet1986,Blanchet2014,Maggiore:GWs_Vol1}. 

In our work, we will resort to the Blanchet-Damour method, which yields reliable descriptions of the motion and radiation of binary BHs only during their early inspiralling stage. In this framework, the generation problem is investigated, for PN (i.e., weakly self-gravitating, weakly stressed, and slowly moving) sources, by splitting it in two sub-problems and by  employing two approximation techniques: in the exterior region, the solution is found by means of the multipolar-post-Minkowskian (MPM) scheme, which combines a post-Minkowskian (PM)  algorithm and a multipolar decomposition, and it is written as a function of either the so-called source multipole moments or  the canonical multipole moments; in the near zone, the solution is constructed through the PN pattern. Since the two procedures have, for PN sources, an overlapping domain of validity, the solutions are then matched together. This strategy permits to express the multipole moments parametrizing the external MPM field in terms of the properties of the source. 

The state-of-the-art computations for compact binaries with vanishing angular momentum give the gravitational waveform   up to 3.5PN order\footnote{We recall that the qualifier $n$PN refers to a  correction of the order $c^{-2n}$ relatively to the \qm{Newtonian} quadrupole formula, which corresponds in turn to a 2.5PN radiation reaction contribution to the equations of motion.} in the case of  circular orbits \cite{Blanchet2014} (with the 4PN mass-type quadrupole moment  recently calculated  in Ref. \cite{Marchand2020}) and  up to 3PN for eccentric orbits \cite{Ebersold2019,Blanchet2009a}; the GW energy flux for binaries  moving on circular orbits has been obtained at 4.5PN  \cite{Marchand2016}; the dynamics is fully known at the 4PN level \cite{Marchand2018}, and a new methodology, referred to as Tutti Frutti \cite{Bini2019a}, has succeeded in extending the current knowledge in the conservative dynamics of binary systems up to the 6PN order   (although some of the underlying  coefficients are missing) \cite{Bini2020c,Bini2020a}. On the other hand, in the case of spinning compact binary systems,  the effects in the radiative field and the energy flux are such that  the spin-orbit (SO) contribution is known at 4PN \cite{Bohe2013c,Marsat2013c}, whereas the spin-spin (SS) \cite{Bohe2015a,Cho2021} and the spin-spin-spin (SSS) \cite{Marsat2014} at 3PN; the SO, SS, and SSS interactions in the equations of motion have been computed at 3.5PN   \cite{Damour2007a,Marsat2012,Bohe2012,Levi_2016}, 3PN \cite{Bohe2015a,Levi2014,Cho2021}, and 3.5PN level \cite{Marsat2014}, respectively.

GWs theory has been also investigated in the context of alternative theories of gravity \cite{Clifton2011,Capozziello2011}. In particular, 2PN gravitational waveforms generated by  a binary of non-spinning compact objects  have been worked out in massless scalar-tensor models by exploiting both the DIRE approach \cite{Mirshekari2013,Lang2014} and the MPM formalism \cite{Sennett2016}.  On the other hand, leading PN spin-orbit effects in quasicircular nonprecessing compact binaries   coupled  to a light scalar field have been recently computed  \cite{Brax2021}. In addition, GWs propagation has been analyzed for a broad class of scalar-tensor theories  by exploiting the geometrical-optics method \cite{Garoffolo2019}.  GWs have been studied also within string-inspired frameworks, such as: Einstein-Maxwell-dilaton \cite{Rincon2020}, Einstein-dilaton Gauss-Bonnet \cite{Carson2020}, Einstein-Maxwell-dilaton-axion models \cite{Lim2019}. Lastly, in Ref. \cite{Shiralilou2021} analytical waveforms for the inspiral stage of a compact binary system in  scalar Gauss-Bonnet gravity patterns have been constructed up to 1PN order  (further results can be found in Refs. \cite{East2020,Odintsov2020}). 

GWs can also represent a promising tool to test quantum gravity through the spectroscopy analysis of more sensitive data \cite{Camelia1998}.  In the literature, some proposals  have been set forth. Some authors argue that GWs can be used as a probe to reveal or constrain quantum modifications at the horizon scale to BH dynamics and address important issues, like: the BH information loss paradox, the firewall phenomenon, and the gravastar proposal \cite{Giddings2016,Cardoso:2016rao,Mazur:2004fk}.  The  quantum-induced fluctuations  which affect two freely falling bodies in a quantized gravitational field might be measured by GWs observatories \cite{Parikh2020}, providing thus valuable hints regarding the existence of gravitons and relevant information about the properties of the GWs source itself \cite{Parikh:2020fhy}. Such a research program is devoted to the exploration via GWs interferometers of the quantum properties of the spacetime, i.e., its \qm{foamy} or \qm{fuzzy} structure \cite{Camelia1998,Camelia1999}. Furthermore, the uncertainty in the arm length of a detector due to quantum metric  fluctuations, as predicted by the theory of quantum gravity  supplemented by the holographic principle, leads to a signal that could be observed at macroscopic distances by a GW apparatus, as discussed in Refs. \cite{Verlinde2019,Verlinde2019b}. A pivotal role in  multi-messenger astronomy will be assumed by the project  HERMES-SP (High Energy Rapid Modular Ensemble of Satellites - Scientific Pathfinder), which will work  in parallel with  LIGO, Virgo, and KAGRA \cite{Fiore:2021smn}. It has been launched with the purpose of detecting and localizing bright high-energy transients such as gamma-ray bursts,  which have been recognized as the electromagnetic counterparts of GWs events \cite{Monitor:2017mdv} and are  important means to infer precious information  on the quantum aspects of GR  \cite{AmelinoCamelia:1997gz}. Among its various scientific objects, there are indeed tests concerning the  granular structure of the spacetime at the Planck scale \cite{Sanna:2021qwp}. 

Motivated by the abovementioned theoretical and observational proposals investigating the emergence of possible quantum phenomena within GWs realm, in this manuscript we study the GWs generation problem in the context of Einstein-Cartan (EC) theory (also known as $U_4$ or Einstein-Cartan-Sciama-Kibble theory) by exploiting the Blanchet-Damour formalism. EC model provides a complete geometrical description of matter at microscopic level since it predicts that the torsion tensor, defined as the antisymmetric part of the connection, couples to the  spin  of elementary particles, in analogy to the well-known coupling  between the metric and the energy-momentum tensor \cite{Hehl1976_fundations,Gasperini-DeSabbata}. Hereafter, the word \qm{spin} indicates the intrinsic \emph{quantum} angular momentum of particles, in contrast with the \emph{classical} angular momentum due to a macroscopic rotation \cite{Hehl1974}.

The research  carried out in this paper aims at understanding possible quantum imprints in the propagation of GWs produced by PN sources in EC theory, namely spinning, weakly self-gravitating, slowly moving, and weakly stressed sources. The paper is organized as follows. In Sec. \ref{sec:problem} we briefly outline the Blanchet-Damour formalism. In Sec. \ref{sec:GW_EC_theory} we deal with the GWs generation problem at 1PN level in EC theory. Eventually, concluding remarks are made in Sec. \ref{Sec:conclusion}.

\emph{Notations}. We use metric signature  $(-,+,+,+)$. Greek indices take values  $0,1,2,3$, while the Latin ones  $1,2,3$. The flat metric is indicated by $\eta^{\alpha \beta }=\eta_{\alpha \beta }={\rm diag}(-1,1,1,1)$. The determinant of the metric $g_{\mu \nu}$ is denoted with $g$. $\epsilon_{ijk}$ is the standard Levi-Civita symbol with $\epsilon_{123}=1$. Round (resp. square) brackets around a pair of indices stands for the usual symmetrization (resp. antisymmetrization) procedure. 

\section{Gravitational Waves in General Relativity:  Blanchet-Damour approach} \label{sec:problem}
We outline the general relativistic GWs theory by first introducing the mathematical problem (see Sec. \ref{sec:mathpro}) and then its approximate resolution following the approach devised by Blanchet and Damour (see Sec. \ref{sec:resolution}).

\subsection{Mathematical problem of GWs generation}
\label{sec:mathpro}

We consider a spacetime $\mathscr{M}$  endowed with a metric $g_{\mu\nu}$ and define the gravitational field amplitude \cite{Blanchet2014}
\begin{equation}\label{eq:def_of_h_alphabeta}
\mathfrak{h}^{\alpha\beta} \equiv \mathfrak{g}^{\alpha\beta}-\eta^{\alpha\beta},
\end{equation}
where $\mathfrak{g}^{\alpha\beta} \equiv \sqrt{-g}g^{\alpha \beta}$ denotes the inverse gothic metric. 

The GWs generation problem in GR is described  by the following set of equations \cite{Blanchet-Damour1986,Poisson-Will2014,Maggiore:GWs_Vol1,Blanchet2014}:
\begin{subnumcases}{\label{eq:pro}}
 \Box \mathfrak{h}^{\alpha\beta}=\chi\mathfrak{T}^{\alpha\beta}, & \label{eq:proa}\\ 
\partial_\beta\mathfrak{h}^{\alpha\beta}=0, & \label{eq:prob}\\ 
\lim_{|\boldsymbol{x}|\to +\infty}\mathfrak{h}^{\alpha\beta}(t,\boldsymbol{x})=0, & \mbox{for}\ $t\le -\mathcal{T}$ \label{eq:proc},\\ 
\partial_t\mathfrak{h}^{\alpha\beta}(t,\boldsymbol{x}) =0, &  \mbox{for}\ $t\le -\mathcal{T}$, \label{eq:prod} 
 \end{subnumcases}
where $\Box \equiv \eta^{\alpha \beta}\partial_\alpha \partial_\beta$, $\chi \equiv 16\pi G/c^4$, 
\begin{eqnarray} \label{eq:tau-alpha-beta}
\mathfrak{T}^{\alpha\beta}\equiv \left(-g\right) T^{\alpha\beta}+\frac{1}{\chi}\Lambda^{\alpha\beta},
\end{eqnarray}
is the \emph{effective stress-energy pseudo-tensor}, and 
\begin{equation} \label{eq:Lambda-alpha-beta-GR}
 \Lambda^{\alpha\beta} \equiv \chi \left(-g\right) \left(t^{\alpha \beta}_{\rm LL}+t^{\alpha \beta}_{\rm H}\right),   
\end{equation}
 $t^{\alpha \beta}_{\rm LL}$ being the \emph{Landau-Lifshitz pseudo-tensor} \cite{Landau-Lifschitz}, whose expression, worked out in harmonic coordinates, reads as
\begin{align}
\label{eq:tLL-GR}
\chi (-g)t^{\alpha \beta}_{\rm LL} &= \frac{1}{2}g^{\alpha \beta}g_{\gamma \delta}\partial_\epsilon \mathfrak{h}^{\gamma \kappa} \partial_\kappa \mathfrak{h}^{\delta \epsilon} - 2 g_{\delta \epsilon}g^{\gamma (\alpha}\partial_\kappa \mathfrak{h}^{\beta) \epsilon}\partial_\gamma \mathfrak{h}^{\delta \kappa}  \nonumber \\
& + g_{\gamma \delta}g^{\epsilon \kappa}\partial_\epsilon \mathfrak{h}^{\alpha \gamma}\partial_\kappa \mathfrak{h}^{\beta \delta} + \frac{1}{8}\left(2g^{\alpha \gamma}g^{\beta \delta}-g^{\alpha \beta}g^{\gamma \delta}\right)  \nonumber \\
& \times \left(2g_{\epsilon \kappa}g_{\lambda \sigma}-g_{\kappa \lambda}g_{\epsilon \sigma}\right)\partial_\gamma \mathfrak{h}^{\epsilon \sigma}\partial_\delta \mathfrak{h}^{\kappa \lambda},
\end{align}
and  $t^{\alpha \beta}_{\rm H}$   an additional harmonic-gauge contribution  which can be written as
\begin{equation} \label{t-alpha-beta-H-1-GR}
\chi \left(-g\right) t^{\alpha \beta}_{\rm H} =  \partial_\mu \partial_\rho \left( \mathfrak{h}^{\alpha \rho}\mathfrak{h}^{\beta \mu}- \mathfrak{h}^{\alpha \beta} \mathfrak{h}^{\mu \rho}\right).
\end{equation}

The gauge condition (\ref{eq:prob}) is equivalent to the matter equations of motion, i.e.,   \cite{Landau-Lifschitz,Blanchet2014}
\begin{eqnarray} \label{eq:conservation-laws-GR}
\partial_{\beta} \mathfrak{h}^{\alpha\beta} =0\quad \Leftrightarrow\quad \partial_{\beta} \mathfrak{T}^{\alpha\beta} =0\quad \Leftrightarrow\quad \nabla_\beta T^{\alpha \beta}=0.
\end{eqnarray}

\begin{figure}[h!]
    \centering
    \includegraphics[scale=0.3]{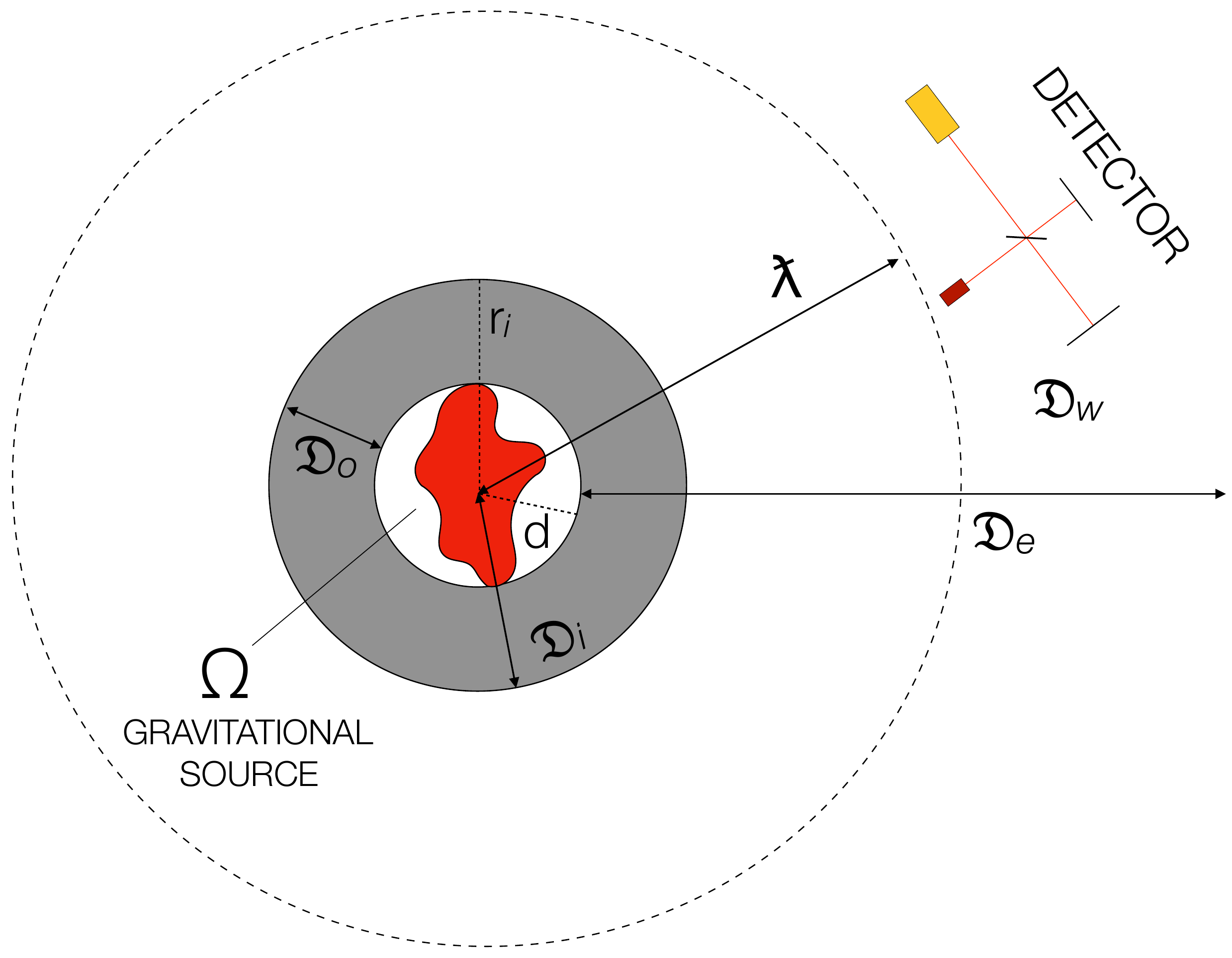}
    \caption{The  domains  in which  the spatial part $\mathbb{R}^3$ of $\mathscr{M}$ is decomposed: gravitational source location $\Omega$, inner zone $\mathfrak{D}_i\supset\Omega$, overlapping zone $\mathfrak{D}_o=\mathfrak{D}_i\cap\mathfrak{D}_e$, exterior zone $\mathfrak{D}_e$. The dashed circle having radius $\lambdabar$ permits to define the wave zone $\mathfrak{D}_w\subset \mathfrak{D}_e$,  the region where a  detector is framed.}
    \label{fig:Fig1}
\end{figure}

The mathematical problem (\ref{eq:pro}) is defined in $\mathscr{M}$, which we assume to be homeomorphic to $\mathbb{R}^4$  and topologically split in space $\mathbb{R}^3$ and time $\mathbb{R}$. Since the gravitational source is localized, $T^{\alpha\beta}\in C^{\infty}_{\rm comp}\left(\mathbb{R}^4,\Omega\right)$, i.e., it is a smooth function in $\mathbb{R}^4$ endowed with the following spatially compact support:
\begin{equation}\label{eq:Omega-region-GR}
\Omega=\left\{\boldsymbol{x}\in\mathbb{R}^3:|\boldsymbol{x}|\leq d\right\}, \end{equation}
where $\vert \boldsymbol{x}\vert \equiv r=\sqrt{\delta_{ij}x^i x^j}$ is the harmonic-coordinate radial distance and $d$ denotes the typical size of the source. Since the source is supposed to be PN, the reduced wavelength $\lambdabar$ of the gravitational radiation  is such that $\lambdabar \gg d$. Therefore, the spatial domain $\mathbb{R}^3$ can be further decomposed as $\mathbb{R}^3=\mathfrak{D}_e\cup\mathfrak{D}_i$, where (see Fig. \ref{fig:Fig1})
\begin{eqnarray}
&&\mathfrak{D}_e=\left\{\boldsymbol{x}\in\mathbb{R}^3\, : \,d<|\boldsymbol{x}|\right\},\label{eq:exterior_Domain}\\
&&\mathfrak{D}_i=\left\{\boldsymbol{x}\in\mathbb{R}^3\, : |\boldsymbol{x}| < r_i,\mbox{\ with}\ d<r_i \ll \lambdabar \right\}. \label{eq:interior_Domain}   
\end{eqnarray}
The set $\mathfrak{D}_i$ is called \emph{interior zone}, while $\mathfrak{D}_e$ \emph{exterior zone}. For a PN source these two zones overlap in the so-called \emph{overlapping  region} $\mathfrak{D}_o$,  being thus defined by \cite{Blanchet2014}
\begin{equation} \label{eq:overlap_Domain}
\mathfrak{D}_o=\mathfrak{D}_e\cap\mathfrak{D}_i =\left\{\boldsymbol{x}\in\mathbb{R}^3\, : \,d<|\boldsymbol{x}|<r_i\right\}.
\end{equation}
Finally, the spatial region where a detector apparatus is located is known as \emph{wave zone}, formally defined as
\begin{equation}
\mathfrak{D}_w=\left\{\boldsymbol{x}\in\mathbb{R}^3\, : \,\lambdabar\ll|\boldsymbol{x}|\right\}.\label{eq:wave_zone}
\end{equation} 

\subsection{Approximate resolution}
\label{sec:resolution}
The fundamental problem of GWs generation consists in formally relating the material content of the  source in $\Omega$  to the asymptotic gravitational wave field $\mathfrak{h}_{\alpha \beta}$ in $\mathfrak{D}_w$ \cite{Blanchet-Damour1986,Blanchet2014}. This issue cannot be solved analytically and   appropriate approximation methods  should be invoked. 

In the exterior domain $\mathfrak{D}_e$, it is supposed that the metric (\ref{eq:def_of_h_alphabeta}) admits the following MPM expansion \cite{Blanchet-Damour1986}: 
\begin{equation}
\label{PM_expansion1}
\mathfrak{h}_{\rm ext}^{\alpha\beta} = \sum_{n=1}^{+\infty} G^n\mathfrak{h}_{(n)}^{\alpha\beta},
\end{equation}
where  each  $\mathfrak{h}_{(n)}^{\alpha\beta}$ admits a \emph{finite multipolar expansion}.  Equation (\ref{PM_expansion1}) is exploited to solve perturbatively in  $\mathfrak{D}_e$ the vacuum Einstein equations and  the resulting MPM solution is parametrized by the symmetric-trace-free (STF) \emph{source multipole moments}, defined as \cite{Blanchet2014,Maggiore:GWs_Vol1}\footnote{We  use the \emph{multi-index notation}, where $L$ denotes the multi-index $i_1i_2\dots i_L$, made of $l$ spatial indices \cite{Blanchet-Damour1986}.   \label{footnote-multindex}}
\begin{equation}
\label{eq:source_multipole_moments}
    \mathfrak{h}_{\rm ext}^{\alpha\beta} = \mathfrak{h}_{\rm ext}^{\alpha\beta} \left(I_L,J_L,W_L,X_L,Y_L,Z_L \right),
\end{equation}
where $I_L$ and $J_L$ are the \emph{mass-moments of order $l$} and \emph{current-moments of order $l$}, respectively, while $W_L,X_L,Y_L,Z_L$ are the \emph{gauge moments of order $l$}. A coordinate transformation allows to define a canonical metric depending on the STF  \emph{canonical mass-type $M_L$} and \emph{current-type $S_L$ moments} \cite{Thorne1980,Blanchet2014}.

By employing  \emph{radiative coordinates}, the MPM solution can be written in terms of the SFT   \emph{mass-type radiative moments} $U_L$ and \emph{current-type radiative moments} $V_L$ \cite{Bondi1962,Sachs1962,Penrose1963,Blanchet-Damour1986,Blanchet1986},  which are related to the canonical ones via some highly non-linear relations \cite{Blanchet2014}.

In the near-zone $\mathfrak{D}_i$, the Einstein field equations are perturbatively solved through the PN iteration \cite{Poujade-Blanchet2002, Blanchet2014}, where the  field amplitude $\mathfrak{h}^{\alpha \beta}$
admits the general  PN expansion (indicated with an overline) 
\begin{equation}
\bar{\mathfrak{h}}^{\alpha\beta} (t,\boldsymbol{x};c)=\sum_{m=2}^{+\infty} \frac{1}{c^m}\,{}^{(m)}\bar{\mathfrak{h}}^{\alpha \beta}(t,\boldsymbol{x};\log c).
\label{eq:PN_expansion}
\end{equation}

The matching procedure allows to obtain the explicit expressions of the source multipole moments and 
stems from the existence of the overlapping region $\mathfrak{D}_o$, where both  MPM and PN expansions are valid. The matching equation  reads  as (spacetime indices suppressed) \cite{Blanchet1998,Blanchet2014}
\begin{equation}
\label{eq:matching equation}
    \overline{\mathcal{M}\left(\mathfrak{h}\right)}= \mathcal{M}\left(\bar{\mathfrak{h}}\right),
\end{equation}
where $\overline{\mathcal{M}\left(\mathfrak{h}\right)}$ is the near-zone  re-expansion of the multipolar expansion of the radiative field, whereas $\mathcal{M}\left(\bar{\mathfrak{h}}\right)$ is  the far-zone re-expansion of the PN expansion  $\bar{\mathfrak{h}}$. 

\section{Gravitational Waves in Einstein-Cartan theory}
\label{sec:GW_EC_theory}
This section concerns GWs emitted by PN sources in the context of EC theory.  A brief account of EC model is given in Sec. \ref{Sec:Einstein-Cartan theory}. This prepares the ground for the analysis of the mathematical problem underlying the GWs generation,  which is presented in Sec. \ref{sec:ori_mat_pro}. The resolution of this problem, when sources are negligibly self-gravitating, can be tackled via 
linearized EC theory, which we discuss in Sec. \ref{sec:Linearized-EC}.  On the other hand, for weakly self-gravitating spinning sources, the problem can be approximately solved by the Blanchet-Damour method, which has been set out in Sec. \ref{sec:problem}. In this case,   upon introducing the domains (\ref{eq:exterior_Domain})--(\ref{eq:wave_zone}) (see Fig. \ref{fig:Fig1}), the solution is built up in three stages: the inner field is examined by resorting to the PN method in the near zone $\mathfrak{D}_i$  (see Secs. \ref{Sec:PN-expansion-EC} and \ref{Sec:Multipole_analysis_inner_torsion}); the gravitational field in the external vacuum region $\mathfrak{D}_e$  is investigated by exploiting the MPM scheme  (see Sec. \ref{Sec:External_Metric_and_Matching}), and finally the two solutions are matched in the overlap domain $\mathfrak{D}_0$ (see Sec. \ref{Sec:Matching-procedure}, where we will see that a matching procedure less general than the one described in Sec. \ref{sec:resolution} is sufficient for our purpose). In Sec. \ref{Sec:The lowest-order source multipole moments} we perform a physical analysis of the non-radiative source multipole moments. The section ends with the evaluation of the 1PN asymptotic waveform, see  Sec. \ref{Sec:asymptotic-waveform}.

\subsection{Einstein-Cartan theory} 
\label{Sec:Einstein-Cartan theory}
Gravitational interaction is the only  fundamental force of nature which is  not fully understood at  microscopic level. For this reason, it seems necessary to extend the principles of GR to the microphysical realm. This objective may be framed within EC model, where the novel  object is represented by the torsion tensor. As we will see, the torsion tensor is  the geometrical counterpart of the  intrinsic spin  carried by particles, in analogy to the GR correspondence  between the curvature of the spacetime and the mass-energy of matter fields.

EC theory is defined on a spacetime $\mathscr{M}$ endowed with a metric tensor $g_{\alpha \beta}$ and with the most general metric-compatible affine connection 
\begin{eqnarray} 
\Gamma^\lambda_{\mu \nu}&=&\hat{\Gamma}^{\lambda}_{\mu \nu}-K_{\mu \nu}^{\phantom{\mu \nu} \lambda},\label{eq:Affine_Connection_1}\\
K_{\mu \nu}^{\phantom{\mu \nu} \lambda}&=&  S_{\nu \phantom{\lambda} \mu}^{\phantom{\nu} \lambda}- S^{\lambda}_{\phantom{\lambda} \mu \nu} - S_{\mu \nu}^{\phantom{\mu \nu} \lambda}=-K_{\mu \phantom{\lambda}\nu}^{\phantom{\mu}\lambda},\label{eq:contorsion_tensor}
\end{eqnarray}
where $\hat{\Gamma}^{\lambda}_{\mu \nu}$ denotes the \emph{Christoffel symbols} and $K_{\mu \nu}^{\phantom{\mu \nu} \lambda}$ the \emph{contortion tensor}. Hereafter, the hat symbol denotes quantities framed in GR.  

The antisymmetric and symmetric parts of  (\ref{eq:Affine_Connection_1}) read as, respectively,
\begin{eqnarray} 
\Gamma^\lambda_{[\mu \nu]}& \equiv& \frac{1}{2}(\Gamma^\lambda_{\mu \nu}-\Gamma^\lambda_{\nu \mu})  \equiv S_{\mu \nu}^{\phantom{\mu \nu} \lambda}, \label{eq:torsion_tensor}\\
\Gamma^\lambda_{(\mu \nu)}&\equiv&\frac{1}{2}(\Gamma^\lambda_{\mu \nu}+\Gamma^\lambda_{\nu \mu})=   \hat{\Gamma}^{\lambda}_{\mu \nu} +2 S^{\lambda}_{\phantom{\lambda} (\mu \nu)},\label{eq:symmetric_part}
\end{eqnarray}
where $S_{\mu \nu}^{\phantom{\mu \nu} \lambda}$ is the \emph{Cartan torsion tensor}. 

The affine connection (\ref{eq:Affine_Connection_1}) permits to define the covariant derivative operator $\nabla$ and the \emph{modified covariant derivative} operator $ \overstarbf{\nabla}$, whose action on a generic tensor field of type $(1,1)$ is given by, respectively, 
\begin{eqnarray}
 \nabla_\nu A^{\alpha}_{\phantom{\alpha}\beta} &=& \partial_\nu  A^{\alpha}_{\phantom{\alpha}\beta} + \Gamma^{\alpha}_{\phantom{\alpha}\nu \mu} A^{\mu}_{\phantom{\mu}\beta}  - \Gamma^{\mu}_{\phantom{\mu}\nu \beta} A^{\alpha}_{\phantom{\alpha}\mu},\\
    \overstarbf{\nabla}_\alpha A^{\mu}_{\phantom{\mu}\nu}&=&\left(\nabla_\alpha+2S_{\alpha\beta}{}^\beta\right)A^{\mu}_{\phantom{\mu}\nu}. \label{eq:modified-divergence}
\end{eqnarray}
The  Riemann tensor 
\begin{equation} \label{eq:EC-Riemann-tensor}
R^{\mu}_{\phantom{\mu}\nu \rho \sigma}= \partial_\rho \Gamma^{\mu}_{\sigma\nu }-\partial_\sigma \Gamma^{\mu}_{\rho \nu} + \Gamma^{\mu}_{\rho \alpha} \Gamma^{\alpha}_{ \sigma\nu}-\Gamma^{\mu}_{\sigma\alpha} \Gamma^{\alpha}_{\rho\nu},
\end{equation}
permits to define the (asymmetric) Ricci tensor $R_{\mu\nu}=R^{\alpha}_{\phantom{\mu}\mu \alpha \nu}$  and the  Ricci curvature $R=R_{\mu}{}^\mu$, which in turn allow to write  the (asymmetric) Einstein tensor as
\begin{equation} \label{eq:definition-Einstein-tensor}
G_{\alpha\beta}=R_{\alpha\beta}-\frac{1}{2}g_{\alpha\beta}R.    
\end{equation}

Given the matter Lagrangian  density $\mathcal{L}_{\rm m} =\mathcal{L}_{\rm m} \left(\Psi,\nabla\Psi,g\right)$, supposed to be minimally coupled to the gravitational field \cite{Hehl1976_fundations} and with $\Psi$ denoting a generic matter field, it is possible to introduce the metric energy-momentum tensor $T^{\alpha \beta}$, the  spin angular momentum tensor $\tau_{\gamma}^{\phantom{\gamma}\beta \alpha}$, and the spin energy potential $\mu_{\gamma}^{\phantom{\gamma}\beta \alpha}$ as \cite{Hehl1973b,Hehl1976_fundations}
\begin{eqnarray}
T^{\alpha\beta}&=&\frac{2}{\sqrt{-g}}\frac{\delta\mathcal{L}_{\rm m}}{\delta g_{\alpha\beta}},  \label{eq:metric_stress-energy_tensor} \\
 \tau_{\gamma}^{\phantom{\gamma}\beta \alpha}&=&\frac{1}{\sqrt{-g}}\frac{\delta \mathcal{L}_{\rm m}}{\delta K_{\alpha\beta}{}^\gamma},  \label{eq:canonical_spin_tensor} \\
    \mu_{\gamma}^{\phantom{\gamma}\beta \alpha}&=&\frac{1}{\sqrt{-g}}\frac{\delta \mathcal{L}_{\rm m}}{\delta S_{\alpha\beta}{}^\gamma}. \label{eq:spin_energy_potential}
\end{eqnarray}
Furthermore, by employing Eq. (\ref{eq:contorsion_tensor}),  the tensors $\mu_{\gamma}^{\phantom{\gamma}\beta \alpha}$ and $\tau_{\gamma}^{\phantom{\gamma}\beta \alpha}$ can be related in the following way:  
\begin{equation} \label{eq:mu_tensor}
 \mu^{\alpha\beta\gamma}=-\tau^{\alpha\beta\gamma}+\tau^{\beta\gamma\alpha}-\tau^{\gamma\alpha\beta}.
\end{equation}

The variation of $\mathcal{L}_{\rm m}$ with respect to the metric and the torsion, Eqs. (\ref{eq:metric_stress-energy_tensor}) and (\ref{eq:spin_energy_potential}), contributes to the total energy of  matter.  The total energy-momentum tensor of matter $\mathbb{T}^{\alpha\beta}$ can be defined as  \cite{Hehl1974b,Hehl1976_fundations}
\begin{equation}
\label{eq:canonical_stress-energy_tensor}
\begin{aligned}
\mathbb{T}^{\alpha\beta}&=T^{\alpha\beta}-\overstarbf{\nabla}_\gamma\left( \mu^{\alpha\beta\gamma}\right),
\end{aligned}
\end{equation}
and it is known   to coincide with the   canonical  energy-momentum tensor, whose expression is \cite{Hehl1974b,Hehl1976_fundations,Gasperini-DeSabbata,Hehl1974}
\begin{equation} \label{eq:definition-of-canonical-stress-energy-tensor}
    \mathbb{T}^{\phantom{\beta}\alpha}_{\beta}= \dfrac{1}{\sqrt{-g}}\left[\delta^{\;\alpha}_{\beta} \mathcal{L}_{\rm m}-   \dfrac{\partial \mathcal{L}_{\rm m}}{\partial\left(\partial_\alpha \Psi \right)} \nabla_\beta \Psi\right].
\end{equation}

EC field equations can be written as (recalling that $\chi \equiv 16 \pi G/c^4$)
\begin{subequations} \label{eq:Einstein-Cartan_equations}
\begin{align} 
\hat{G}^{\alpha\beta}&=\frac{\chi}{2}\Theta^{\alpha\beta},\label{eq:hat-G-equals-tilde-T}\\
\Theta^{\alpha\beta}&=T^{\alpha\beta}+\frac{\chi}{2}\mathcal{S}^{\alpha \beta},\label{eq:tilde-T-alpha-beta}    
\end{align}
\end{subequations}
where the tensor $\mathcal{S}^{\alpha \beta}$ reads as \cite{Hehl1974,Hehl1974b,Hehl1976_fundations}
\begin{align}
\label{Eq:S_alphabeta}
    \mathcal{S}^{\alpha \beta} & \equiv
    -4\tau^{\alpha\gamma}{}_{[\delta}\tau^{\beta\delta}{}_{\gamma]}-2\tau^{\alpha\gamma\delta}\tau^\beta{}_{\gamma\delta}
    \nonumber \\
    &+\tau^{\gamma\delta\alpha}\tau_{\gamma\delta}{}^\beta+\frac{1}{2}g^{\alpha\beta}\left(
    4\tau_{\mu}^{\phantom{\mu}\gamma}{}_{\phantom{\gamma}[\delta}\;\tau^{\mu\delta}{}_{\gamma]}+\tau^{\mu\gamma\delta}\tau_{\mu\gamma\delta}\right),
\end{align}
and $\hat{G}^{\alpha\beta}$ is the Riemannian part of the Einstein tensor (\ref{eq:definition-Einstein-tensor}). In  Eq. (\ref{eq:Einstein-Cartan_equations}) we have formally removed the torsion, since we have employed the following identities \cite{Gasperini-DeSabbata}: 
\begin{eqnarray} 
    S_{\mu \nu}^{\phantom{\mu \nu}\lambda} &=& \dfrac{\chi}{2} \left(\tau_{\mu \nu}^{\phantom{\mu \nu}\lambda} + \delta^\lambda_{[\mu} \tau_{\nu]\rho}^{\phantom{\nu]\rho}\rho}\right),\label{eq:torsion&tau} \\
      K_{\mu \nu}^{\phantom{\mu \nu}\alpha}&=&\dfrac{\chi}{2} \left(-\tau_{\mu \nu}^{\phantom{\mu \nu}\alpha} +\tau_{\nu \phantom{\alpha}\mu}^{\phantom{\nu}\alpha}-\tau^\alpha_{\phantom{\alpha}\mu \nu}\right. \label{eq:contorsion&tau}  \\
    &&\left.-\delta^\alpha_\mu \tau_{\nu \rho}^{\phantom{\nu \rho}\rho} +g_{\mu \nu} \tau^{\alpha \phantom{\rho}\rho}_{\phantom{\rho} \rho} \right). \nonumber
\end{eqnarray}
The combined energy-momentum tensor $\Theta^{\alpha\beta}$ occurring in Eq. (\ref{eq:Einstein-Cartan_equations}) contains spin contributions implicitly in $T^{\alpha \beta}$ and explicitly in $\mathcal{S}^{\alpha \beta}$. Moreover, it satisfies the relation 
\begin{equation} \label{eq:nabla_tilde_T}
    \hat{\nabla}_\beta \Theta^{\alpha\beta}=0,
\end{equation}
where $\hat{\nabla}_\beta$ denotes henceforth the covariant derivative with respect to the Levi-Civita connection. 

In EC theory, the dynamical equations of the matter source can be obtained by means of the (generalized) conservation laws of energy-momentum and  angular momentum, which read as, respectively, \cite{Hehl1974b,Hehl1976_fundations,Gasperini-DeSabbata}
\begin{eqnarray}  
\overstarbf{\nabla}_\nu  \mathbb{T}_{\mu}^{\phantom{\mu}\nu} &=& 2 \mathbb{T}_{\lambda}^{\phantom{\lambda}\nu} S_{\mu \nu}^{\phantom{\mu \nu }\lambda} -  \tau_{\nu \rho}^{\phantom{\nu \rho}\sigma} R_{\mu \sigma}^{\phantom{\mu \sigma}\nu \rho},\label{eq:conservation-law-energy-momentum}\\
\overstarbf{\nabla}_\lambda \tau_{\mu \nu}^{\phantom{\mu \nu}\lambda} &=& \mathbb{T}_{[\mu \nu]}.\label{eq:conservation-law-angular-momentum}
\end{eqnarray}
In particular, the translational dynamics of a spinning test particle   can be derived from the conservation law (\ref{eq:conservation-law-energy-momentum}) by employing the so-called \qm{pole-particle} approximation \cite{Hehl1971} (see also Ref. \cite{Gasperini-DeSabbata} for further details). In this way, we get the  \emph{Mathisson-Papapetrou-like equations} 
\begin{equation}
     \dfrac{{\rm d}P^\mu}{{\rm d}\tau} + \hat{\Gamma}^{\mu}_{\nu \lambda} P^{(\lambda} u^{\nu)} +  K^{\mu}_{\phantom{\mu} \nu \lambda} P^{[\lambda} u^{\nu]}= - R^{\mu}_{\phantom{\mu}\nu \rho \sigma} S^{\rho \sigma} u^{\nu},
    \label{eq:Mathisson-Papapertrou_eq}
\end{equation}
$\tau$ being the proper time, $u^{\nu}$ the four-velocity, and
\begin{align}
 \label{eq:Mathisson-Papapertrou_parameters}
         P^\mu u^\nu &= u^0 \int {\rm d}^3 \boldsymbol{x}\, \sqrt{-g} \, \dfrac{\mathbb{T}^{\mu \nu}}{c}, 
         \\
         S^{\mu \nu} u^\lambda &= u^0 \int {\rm d}^3 \boldsymbol{x} \, \sqrt{-g}\,  \dfrac{\tau^{\mu \nu \lambda}}{c}, 
         \label{eq:Mathisson-Papapertrou_parameters-spin}
\end{align}
the total momentum and spin angular momentum of the test particle, respectively.

\subsection{Mathematical problem}
\label{sec:ori_mat_pro}

Along the same lines as for GR, we introduce also in EC theory  the gravitational field amplitude (cf. Eq. (\ref{eq:def_of_h_alphabeta})) 
\begin{equation}\label{eq:def_of_h_alphabeta2}
\mathfrak{h}^{\alpha\beta} \equiv \mathfrak{g}^{\alpha\beta}-\eta^{\alpha\beta},
\end{equation}
where $\mathfrak{g}^{\alpha\beta} \equiv \sqrt{-g}g^{\alpha \beta}$ denotes the inverse gothic metric.

The starting point of GWs generation problem is now represented by the EC field equations (\ref{eq:Einstein-Cartan_equations}), which, likewise in GR,  can be written in the form of inhomogeneous flat-space d'Alembertian equations once a generalized harmonic gauge is invoked (see Eq. (\ref{eq:prob-EC})). The differential problem is defined in the spacetime $\mathscr{M}$ which, as in GR, we suppose to be homeomorphic to $\mathbb{R}^4$, and is supplemented by  two boundary conditions already exploited in GR, i.e., Eqs. (\ref{eq:proc}) and (\ref{eq:prod}). In this way, we obtain the following well-posed mathematical problem:
\begin{subnumcases}{\label{eq:math-problem-EC}}
\Box \mathfrak{h}^{\mu\nu}=\chi \left(-g\right)\Theta^{\mu \nu}+\tilde{\Lambda}^{\mu \nu},  & \label{eq:proa-EC}\\ 
H^\alpha=0, & \label{eq:prob-EC}\\ 
\lim \limits_{|\boldsymbol{x}|\to +\infty}\mathfrak{h}^{\alpha\beta}(t,\boldsymbol{x})=0, \qquad \mbox{for}\ t\le -\mathcal{T}, & \label{eq:proc-EC}\\ 
\partial_t\mathfrak{h}^{\alpha\beta}(t,\boldsymbol{x}) =0,\hspace{1.55cm} \mbox{for}\ t\le -\mathcal{T}.  & \label{eq:prod-EC} 
 \end{subnumcases}

If we define $\Box_g \equiv g^{\mu\nu}\nabla_\mu\nabla_\nu$, and $\hat{\Box}_g \equiv g^{\mu\nu}\hat{\nabla}_\mu\hat{\nabla}_\nu$, the generalized gauge condition (\ref{eq:prob-EC}) reads as
\begin{eqnarray}
    H^\alpha &\equiv&\Box_g x^\alpha = \hat{H}^\alpha - 2 S^{\alpha \mu}_{\phantom{\alpha \mu} \mu} =0, \label{eq:H-alpha-EC}
    \\
    \hat{H}^\alpha &\equiv& \hat{\Box}_g x^\alpha = \dfrac{1}{\sqrt{-g}} \partial_\lambda \mathfrak{h}^{\alpha \lambda},
    \label{eq:H-hat-alpha-EC}
\end{eqnarray}
and hence it is equivalent to
\begin{equation} \label{eq:gauge-harmonic-h-alpha-beta-EC}
    \partial_\lambda \mathfrak{h}^{\alpha \lambda} = 2 \sqrt{-g} S^{\alpha \mu}_{\phantom{\alpha \mu} \mu}.
\end{equation}
It is clear that due to the presence of the torsion tensor, we have obtained a gauge condition which differs from the corresponding expression  employed in GR (cf. Eq. (\ref{eq:prob})). For this reason,  the right-hand side of EC equations (\ref{eq:proa-EC}) will exhibit new contributions, which generalize  those occurring in Einstein equations.  These new terms are included  in the EC pseudo-tensor $\tilde{\Lambda}^{\mu \nu}$, which can be written as (cf. Eq. (\ref{eq:Lambda-alpha-beta-GR}))
\begin{equation} \label{tilde-Lambda-EC}
    \tilde{\Lambda}^{\mu \nu} = \chi \left(-g\right) \left(\tilde{t}^{\mu \nu}_{\rm LL}+\tilde{t}^{\mu \nu}_{\rm H}+t^{\mu \nu}_{\rm S}\right),
\end{equation}
where $\tilde{t}^{\mu \nu}_{\rm LL}$ and $\tilde{t}^{\mu \nu}_{\rm H}$ are the generalized version of 
the Landau-Lifshitz pseudo-tensor (\ref{eq:tLL-GR}) and of Eq. (\ref{t-alpha-beta-H-1-GR}), respectively,  while  $t^{\mu \nu}_{\rm S}$ is a new additional contribution due entirely to the torsion (and for this reason we have not  indicated it with a tilde).  Their expressions are:
\begin{eqnarray}   
\chi \left(-g\right)\tilde{t}^{\mu \nu}_{\rm LL} &=& \chi \left(-g\right)t^{\mu \nu}_{\rm LL} 
 \nonumber \\ 
& +& \left(2 \sqrt{-g} S^{\gamma \epsilon}_{\phantom{\gamma \epsilon} \epsilon}\right)\partial_\gamma \mathfrak{h}^{\mu \nu}-4\left(-g\right) S^{\mu \epsilon}_{\phantom{\mu \epsilon}\epsilon} S^{\nu \tau}_{\phantom{\nu \tau}\tau}, \label{eq:tilde-t-LL-EC}
\\
\chi \left(-g\right) \tilde{t}^{\mu \nu}_{\rm H} &=&  \chi \left(-g\right) t^{\mu \nu}_{\rm H} 
 - 4 \left(-g\right) S^{\mu \epsilon}_{\phantom{ \mu \epsilon} \epsilon} S^{\nu \tau}_{\phantom{ \nu \tau } \tau} 
 \nonumber  \\ 
&-& 2 \mathfrak{h}^{\alpha (\nu} \partial_\alpha \left[2\sqrt{-g}S^{\mu) \epsilon}_{\phantom{ \mu) \epsilon} \epsilon}\right]  + \partial_\alpha \left[\mathfrak{h}^{\mu \nu}  \left(2\sqrt{-g}S^{\alpha \epsilon}_{\phantom{ \alpha \epsilon} \epsilon}\right)\right] 
 \nonumber  \\ 
&+&    \left(2\sqrt{-g}S^{\alpha \epsilon}_{\phantom{ \rho \epsilon} \epsilon}\right) \partial_\alpha \mathfrak{h}^{\mu \nu}, \label{eq:tilde-t-H-EC}
  \\
 \chi \left(-g\right) t^{\mu \nu}_{\rm S} &=& -\left(\eta^{\mu \nu}  +\mathfrak{h}^{\mu \nu}\right) \partial_\alpha \left(2 \sqrt{-g}   S^{\alpha \epsilon}_{\phantom{ \alpha \epsilon} \epsilon}\right)  
 \nonumber \\
 & +& 2 \left[\eta^{\alpha (\nu}  +\mathfrak{h}^{\alpha (\nu}\right] \partial_\alpha \left[2 \sqrt{-g}   S^{\mu) \epsilon}_{\phantom{ \mu) \epsilon} \epsilon}\right] 
 \nonumber \\ 
 & +& 4 \left(-g\right) S^{\mu \epsilon}_{\phantom{ \mu \epsilon}  \epsilon} S^{\nu \tau}_{\phantom{ \nu \tau}  \tau} - \left(4 \sqrt{-g} S^{\alpha \epsilon}_{\phantom{ \alpha \epsilon}  \epsilon}\right) \partial_\alpha \mathfrak{h}^{\mu \nu}. \label{eq:-t-S-EC}
\end{eqnarray}
The generalized gauge condition (\ref{eq:prob-EC}) leads to the following identity:
\begin{equation} \label{eq:EC-theory-gauge-cond_1}
    \partial_\nu \left[\chi \left(-g\right)\Theta^{\mu \nu}+\tilde{\Lambda}^{\mu \nu}\right] = \Box \left(2 \sqrt{-g} S^{\mu \epsilon}_{\phantom{ \mu \epsilon}  \epsilon} \right),
\end{equation}
which in turn is equivalent to (\ref{eq:nabla_tilde_T}). In deriving Eq. (\ref{eq:EC-theory-gauge-cond_1}), we have taken into account that
\begin{equation} \label{eq:derivative-tilde-T-tilde-LL-EC}
  \partial_\nu \left[\chi \left(-g\right)\left(\Theta^{\mu \nu}+\tilde{t}^{\mu \nu}_{\rm LL}\right)\right] =0, 
\end{equation}
along with  the following equations: 
\begin{eqnarray}
\partial_\nu \left[\chi \left(-g\right)\tilde{t}^{\mu \nu}_{\rm H}\right] & =& -4 \partial_\nu\left[ \left(-g\right) S^{\mu \epsilon}_{\phantom{ \mu \epsilon}  \epsilon} S^{\nu \tau}_{\phantom{ \nu \tau}  \tau} \right] 
\nonumber \\
 & - & \mathfrak{h}^{\alpha \nu} \partial_\alpha \partial_\nu \left(2 \sqrt{-g}   S^{\mu \epsilon}_{\phantom{ \mu \epsilon} \epsilon}\right) 
 \nonumber \\
 & + &\partial_\alpha \left[ \left(2 \sqrt{-g}   S^{\mu \epsilon}_{\phantom{ \mu \epsilon} \epsilon}\right)  \left(2 \sqrt{-g}   S^{\alpha \tau}_{\phantom{ \alpha \tau} \tau}\right)\right] 
 \nonumber \\
 & +& \left(\partial_\alpha \mathfrak{h}^{\mu \nu}\right)  \partial_\nu \left(2 \sqrt{-g}   S^{\alpha \epsilon}_{\phantom{ \alpha \epsilon} \epsilon}\right),
 \\
 \partial_\nu \left[\chi \left(-g\right)t^{\mu \nu}_{\rm S}\right] & =&  \Box \left(2 \sqrt{-g} S^{\mu \epsilon}_{\phantom{ \mu \epsilon}  \epsilon} \right)
  \nonumber \\
 &-&  \partial_\nu \left[\chi \left(-g\right)\tilde{t}^{\mu \nu}_{\rm H}\right].
 \label{eq:derivative-t-S-EC}
 \end{eqnarray}

\subsubsection{Simplification of the generalized harmonic gauge condition}

In EC theory, the generalized harmonic gauge  (\ref{eq:prob-EC}) amounts to require that the spacetime coordinates $x^\alpha$ satisfy a set of  Riemannian d'Alembertian equations  having a non-vanishing source term due to the presence of $ S^{\alpha \mu}_{\phantom{ \alpha \mu}  \mu}$, see Eqs. (\ref{eq:H-alpha-EC}) and (\ref{eq:H-hat-alpha-EC}). The same factor  also occurs  in the definition of the tensors (\ref{eq:tilde-t-LL-EC})--(\ref{eq:-t-S-EC})  and can be interpreted as a correction to the analogous GR quantities induced by the torsion. Therefore, we can  simplify all the above expressions if we require that
\begin{equation} \label{eq:Frenkel-condition_general-case}
    S^{\alpha \mu}_{\phantom{ \alpha \mu}  \mu}=0.
\end{equation} 
In general, the torsion tensor has 24 independent components, which have been lowered to 20 thanks to the condition (\ref{eq:Frenkel-condition_general-case}). This permits to further reduce the degrees of freedom of the EC theory with respect to the GR case and hence the complexity of the involved computations.

There exist various  physical systems where Eq. (\ref{eq:Frenkel-condition_general-case}) is satisfied. One example is furnished by  the  semiclassical model of the neutral Weyssenhoff spinning fluid, where such an identity follows, via Eq. (\ref{eq:torsion&tau}), from the so-called \emph{Frenkel condition}  \cite{Hehl1976_fundations,Obukhov1987}
\begin{equation}
\tau_{\alpha\beta}{}^\beta \equiv s_{\alpha\beta}u^\beta=0,
\end{equation}
$u^\alpha$ being the fluid velocity  and $s_{\alpha\beta}=s_{[\alpha\beta]}$ the spin density. 
However, there is a plethora of other physical examples where the Frenkel condition is applied: model of  Frenkel spinning electrons and all spinning bodies in curved spacetimes \cite{Ramirez2014}, hydrodynamics of fluids endowed with spin (although these are framed in GR, the spinning proprieties can be suitably recovered also by resorting to the EC theory) \cite{Becattini2011}. Finally, Eq. (\ref{eq:Frenkel-condition_general-case}) is valid whenever the torsion tensor is  totally antisymmetric,  as occurs  for example in the case of the  Dirac field \cite{Hehl1976_fundations}.

Through the assumption (\ref{eq:Frenkel-condition_general-case}), Eq. (\ref{eq:gauge-harmonic-h-alpha-beta-EC}) becomes identical to the GR gauge condition (\ref{eq:prob}). Furthermore, Eqs. (\ref{eq:tilde-t-LL-EC})--(\ref{eq:-t-S-EC}) turn out to be less complex, because $\tilde{t}^{\mu \nu}_{\rm LL}$ and $\tilde{t}^{\mu \nu}_{\rm H}$ reduce to their corresponding GR counterparts (\ref{eq:tLL-GR}) and (\ref{t-alpha-beta-H-1-GR}), respectively, whereas $t^{\alpha \beta}_{\rm S}$ vanishes identically. As a consequence, the gravitational  source term (\ref{tilde-Lambda-EC}) assumes the same form as the GR expression  (\ref{eq:Lambda-alpha-beta-GR}). Therefore, once Eq. (\ref{eq:Frenkel-condition_general-case}) is employed, the mathematical problem (\ref{eq:math-problem-EC}) reads as 
\begin{subnumcases}{\label{eq:math-problem-EC2}}
\Box \mathfrak{h}^{\mu\nu}=\chi \tilde{\mathfrak{T}}^{\mu \nu}, & \label{eq:proa-EC2}\\ 
\partial_\lambda \mathfrak{h}^{\alpha \lambda}=0, & \label{eq:prob-EC2}\\ 
\lim \limits_{|\boldsymbol{x}|\to +\infty}\mathfrak{h}^{\alpha\beta}(t,\boldsymbol{x})=0, \qquad \mbox{for}\ t\le -\mathcal{T}, & \label{eq:proc-EC2}\\ 
\partial_t\mathfrak{h}^{\alpha\beta}(t,\boldsymbol{x}) =0,\hspace{1.55cm} \mbox{for}\ t\le -\mathcal{T}, & \label{eq:prod-EC2} 
 \end{subnumcases}
 where we have defined the EC pseudo-tensor
 \begin{equation} \label{eq:tilde-tau-alpha-beta-def-1}
     \tilde{\mathfrak{T}}^{\alpha\beta}\equiv \left(-g\right)\, \Theta^{\alpha\beta}+\frac{1}{\chi}\Lambda^{\alpha\beta}.
 \end{equation}
The mathematical problem, written as in Eq. (\ref{eq:math-problem-EC2}), resembles that of GR  (see Eqs. (\ref{eq:pro}) and (\ref{eq:tau-alpha-beta})). Furthermore,  bearing in  mind Eqs. (\ref{eq:EC-theory-gauge-cond_1})--(\ref{eq:derivative-t-S-EC}), the condition (\ref{eq:Frenkel-condition_general-case}) allows us to write the following conservation laws:
\begin{eqnarray} \label{eq:conservation-laws-EC}
\partial_{\beta} \mathfrak{h}^{\alpha\beta} =0 \; \Leftrightarrow \; \partial_{\beta} \tilde{\mathfrak{T}}^{\alpha\beta} =0 \; \Leftrightarrow \; \hat{\nabla}_\beta \Theta^{\alpha \beta}=0,
\end{eqnarray}
which turn out to be very similar to Eq. (\ref{eq:conservation-laws-GR}). Subject to the hypothesis (\ref{eq:prod-EC2}), Eq. (\ref{eq:proa-EC2}) can be \emph{formally} recast in the following integro-differential form:
\begin{equation} 
\mathfrak{h}^{\alpha\beta}=\chi \,\Box^{-1}_{\rm ret}\tilde{\mathfrak{T}}^{\alpha\beta},
\end{equation}
where $\Box^{-1}_{\rm ret}$ is the \emph{retarded Green function}, defined as 
\begin{equation}
\label{eq:standard_retarded}
    \Box^{-1}_{\rm ret}\tilde{\mathfrak{T}}^{\alpha\beta}\equiv\frac{-1}{4\pi} \int \limits_{\mathbb{R}^3} \dfrac{{\rm d}^{3}\boldsymbol{x}^{\prime}}{\vert \boldsymbol{x}-\boldsymbol{x}^{\prime}\vert} \tilde{\mathfrak{T}}^{\alpha \beta}\left(t-\dfrac{\vert \boldsymbol{x}-\boldsymbol{x}^{\prime}\vert}{c},\boldsymbol{x}^{\prime}\right).
\end{equation}

In our analysis, we will suppose that the gravitational source is confined to the region $\Omega$, defined by  Eq. (\ref{eq:Omega-region-GR}). This means that for the metric energy-momentum tensor (\ref{eq:metric_stress-energy_tensor}) and  the  spin angular momentum tensor (\ref{eq:canonical_spin_tensor}) we can write $T^{ \alpha \beta} \in C^{\infty}_{\rm comp}(\mathbb{R}^4,\Omega)$ and  $\tau_{\gamma}^{\phantom{\gamma}\beta \alpha} \in C^{\infty}_{\rm comp}(\mathbb{R}^4,\Omega)$. In other words, the combined energy-momentum tensor $\Theta^{\alpha\beta}$ is such that $\Theta^{\alpha\beta}\in C^{\infty}_{\rm comp}(\mathbb{R}^4,\Omega)$ (see Eqs. (\ref{eq:tilde-T-alpha-beta}) and (\ref{Eq:S_alphabeta})). Moreover, the compactness property of $\tau_{\gamma}^{\phantom{\gamma}\beta \alpha}$, jointly with Eq. (\ref{eq:torsion&tau}), permits to conclude that  the torsion tensor will vanish  in the region  $\Omega_{\rm ext} \equiv \mathbb{R}^3 \,\setminus\, \Omega $.  

\subsubsection{Implications of the hypothesis $S^{\alpha \mu}{}_{\mu}=0$}

We  note that if we had not considered the assumption (\ref{eq:Frenkel-condition_general-case}), Eqs. (\ref{eq:H-alpha-EC})--(\ref{eq:gauge-harmonic-h-alpha-beta-EC}) would have led to two different gauge conditions, depending on whether the mathematical problem (\ref{eq:math-problem-EC}) had been addressed in  $\Omega$ (where in general $S^{\alpha \mu}_{\phantom{ \alpha \mu}  \mu} \neq 0$) or in $\Omega_{\rm ext}$ (where the compactness hypothesis  guarantees that $S^{\alpha \mu}_{\phantom{ \alpha \mu}  \mu} =0 $).  On the other hand, having enforced Eq. (\ref{eq:Frenkel-condition_general-case}),  the gauge condition  is the same both in $\Omega$ and $\Omega_{\rm ext}$, where it is simply given by (\ref{eq:prob-EC2}). 
In addition, in the mathematical problem (\ref{eq:math-problem-EC}), the
generalized harmonic gauge defines one set of coordinates $x_{\rm i}$ in  $\Omega$  and another set $x_{\rm e}$ in the external region $\Omega_{\rm ext}$, the former satisfying $\hat{\Box}_g x^\alpha_{\rm i}=2 S^{\alpha \mu}_{\phantom{\alpha \mu} \mu}$ and the latter the harmonic gauge $\hat{\Box}_g x^\alpha_{\rm e}=0$. Since $x_{\rm i}$ and $x_{\rm e}$ are solutions of the inhomogeneous and homogeneous wave equations, respectively, they  are related by the  formula $x_{\rm i}=x_{\rm e}+x_{\rm p}$, where $x_{\rm p}$   is a particular solution of the inhomogeneous equation. Therefore, the generic transformation which permits to move from the inner coordinate system $x_{\rm i}$ to the outer coordinate system $x_{\rm e}$ is the following linear function (defined up to a global multiplication constant): $x_{\rm e}^\mu=f^\mu(x_{\rm i})\equiv x_{\rm i}^\mu-x_{\rm p}^\mu$, while its inverse is given by $x_{\rm i}^\mu=(f^{-1})^\mu(x_{\rm e})\equiv x_{\rm e}^\mu+x_{\rm p}^\mu$.

\subsection{The linearized Einstein-Cartan theory} \label{sec:Linearized-EC}

One way to handle the mathematical problem (\ref{eq:math-problem-EC2}) consists in exploiting  the linearized EC theory, which deals with the regime of weak  metric  and torsion fields \cite{Arkuszewski1974}. We suppose that in the spacetime $\mathscr{M}$  there exists a reference frame in which we can  decompose the metric tensor $g_{\alpha \beta}$ into Minkowski metric $\eta_{\alpha \beta}$ plus a small perturbation
$h_{\alpha \beta}$, i.e., \cite{Romano2019} 
\begin{align} \label{eq:linearizedEC-metric}
g_{\alpha \beta} &= \eta_{\alpha \beta} + h_{\alpha \beta}, \quad \vert h_{\alpha \beta}\vert \ll 1.
\end{align}

The functions $h_{\alpha \beta}$ are related to the gravitational field amplitude  (\ref{eq:def_of_h_alphabeta2}) through 
\begin{equation}
    \mathfrak{h}_{\alpha \beta} = -\theta_{\alpha \beta} + {\rm O}(h^2),
\end{equation}
where we have defined
\begin{equation}
    \theta_{\alpha \beta} \equiv h_{\alpha \beta}- \dfrac{1}{2} \eta_{\alpha \beta}h,
\end{equation}
and  $h \equiv \eta^{\mu \nu}h_{\mu \nu}$ (in linearized theory indices are raised and lowered by $\eta_{\mu \nu}$). Neglecting terms quadratic in $h_{\mu \nu}$,  the \emph{de Donder gauge} (\ref{eq:prob-EC2})  assumes the form 
\begin{equation}\label{eq:linearized_de_Donder_gauge}
    \partial_\nu \theta^{\mu \nu}=0,
\end{equation}
and Eq. (\ref{eq:proc-EC2}) is equivalent to 
\begin{align} 
\lim_{r \to +\infty} h_{\mu \nu}=0 \quad \mbox{for}\ t\le -\mathcal{T},
\end{align}
where the radial distance $r$ is calculated  in terms of coordinates for which Eq. (\ref{eq:linearized_de_Donder_gauge}) is valid. Furthermore, from Eq.  (\ref{eq:prod-EC2}) we obtain, at linear order, 
\begin{equation}
    \partial_t h_{\alpha\beta}=0, \quad \mbox{for}\ t\le -\mathcal{T}.
\end{equation}

If we exploit Eq. (\ref{eq:canonical_stress-energy_tensor}) jointly with Eqs. (\ref{eq:Affine_Connection_1}), (\ref{eq:modified-divergence}), (\ref{eq:torsion&tau}) and (\ref{eq:contorsion&tau}), then by retaining in Eq.  (\ref{eq:proa-EC2}) only those terms linear in both the metric  perturbations  and  the torsion field we obtain 
\begin{equation} \label{eq:linearized_Einstein-Cartan_compact}
         \Box \theta_{\alpha \beta}=- \chi \, T_{\alpha \beta},
\end{equation}
where 
\begin{equation}\label{eq:linearized_U4_Belinfante}
   T_{\alpha \beta} =  \mathbb{T}_{\alpha \beta}+\partial_\gamma \mu_{\alpha \beta }^{\phantom{\alpha \beta} \gamma},
\end{equation}
is the symmetric stress-energy tensor deduced from the linearization  of  Eq. (\ref{eq:canonical_stress-energy_tensor}). The gauge condition (\ref{eq:linearized_de_Donder_gauge}) is equivalent to the flat-space conservation laws for $T^{\alpha \beta}$ and $\mathbb{T}^{\alpha \beta}$, i.e., 
\begin{align}
\partial_\beta \theta^{\alpha \beta}=0 \quad \Leftrightarrow \quad \partial_\beta T^{\alpha \beta}=0 \quad \Leftrightarrow \quad \partial_\beta \mathbb{T}^{\alpha \beta}=0.
  \end{align}

In linearized theory, an important role is fulfilled by the infinitesimal local coordinate transformation  
\begin{equation} \label{eq:linearized-gauge-transformations}
    x^\mu \rightarrow x^{\prime \mu}= x^\mu - \xi^\mu(x),
\end{equation}
$\xi^\mu(x)$ denoting four arbitrary functions small enough to guarantee that $\vert h_{\mu \nu}\vert \ll 1$. Equation (\ref{eq:linearized-gauge-transformations}) induces the following gauge transformation on $h_{\mu \nu}$ \cite{Carroll}:
\begin{equation} \label{eq:transformation-h-Lie}
h_{\mu \nu}(x) \rightarrow h_{\mu \nu}^{\prime}(x^\prime)= h_{\mu \nu}(x) + \left(\pounds_{\xi} \eta\right)_{\mu \nu},
\end{equation}
where (the components of) the  Lie derivative  of the flat  metric  along the vector $\xi^\mu$ is given by \cite{Nakahara2003,Watanabe2004}
\begin{align}
   \left(\pounds_{\xi} \eta\right)_{\mu \nu} &= \xi^\lambda \nabla_\lambda \eta_{\mu \nu} +\eta_{\lambda \nu} \nabla_\mu \xi^\lambda + \eta_{\lambda \mu} \nabla_\nu \xi^\lambda 
    \nonumber \\
    & -2 \xi^\sigma \left(\eta_{\rho \mu} S^{\rho}_{\phantom{\rho}\sigma \nu} +   \eta_{\rho \nu} S^{\rho}_{\phantom{\rho}\sigma \mu} \right). 
\end{align}
Since $\vert \xi^\mu \vert \ll 1$ and both the metric perturbation functions and the torsion field are weak, we obtain, at first order,
\begin{equation}
\left(\pounds_{\xi} \eta\right)_{\mu \nu} = \partial_\mu \xi_\nu + \partial_\nu \xi_\mu,
\end{equation}
and hence from Eq. (\ref{eq:transformation-h-Lie}) we end up with
\begin{equation} \label{eq:transf-on-perturb}
    h_{\mu \nu}(x) \rightarrow h_{\mu \nu}^{\prime}(x^\prime)= h_{\mu \nu}(x) +2 \partial_{(\mu} \xi_{\nu)}.
\end{equation}
Therefore, the linearized gauge transformations of EC theory assume the same form as those of linearized GR \cite{MTW,Maggiore:GWs_Vol1}, provided that $\vert \xi^\mu \vert$, $\vert h_{\mu \nu} \vert$, and $\vert S^{\rho}_{\phantom{\rho}\sigma \nu} \vert$ are small. 
  
To within the precision of the linearized theory, the tiny changes induced in the functional forms of all scalar, vector, and tensor fields by the infinitesimal coordinate transformations (\ref{eq:linearized-gauge-transformations}) can be ignored, except in metric, where the small deviations from the flat metric $\eta_{\mu \nu}$  contain all the information about the gravitational field. This means that, in particular, the Riemann tensor, the stress-energy tensor, and the spin density tensor are unaffected by gauge transformations (\ref{eq:transf-on-perturb}). Let us evaluate the invariance of the Riemann tensor, which will be crucial in Sec. \ref{Sec:Matching-procedure}. In EC model, it can be written as  \cite{Schouten1954,Medina2018} 
\begin{align}
    R_{\mu \nu \rho \sigma} &= \hat{R}_{\mu\nu \rho \sigma} + \hat{\nabla}_\sigma K_{\rho \nu \mu}-\hat{\nabla}_\rho K_{\sigma \nu \mu}
 \nonumber \\
 &+ K_{\rho \alpha \mu}K_{\sigma \nu}^{\phantom{\sigma \nu}\alpha}- K_{\sigma \alpha \mu}K_{\rho \nu}^{\phantom{\rho \nu}\alpha},
\end{align}
and hence at first order we obtain 
\begin{equation}\label{eq:linearized_U4_Riemann}
    R^{{\rm lin}}_{\mu \nu \rho \sigma} = \hat{R}^{\rm lin}_{\mu \nu \rho \sigma} + 2 \partial_{[\sigma}K_{\rho]\nu \mu}, 
\end{equation}
where we know from the weak-field limit of GR  that
\begin{equation}
     \hat{R}^{\rm lin}_{\mu \nu \rho \sigma}= \dfrac{1}{2} \left(\partial_\nu \partial_\rho h_{\mu \sigma}+ \partial_\mu \partial_\sigma h_{\nu \rho} -\partial_\mu \partial_\rho h_{\nu \sigma}-\partial_\nu \partial_\sigma h_{\mu \rho}\right).
\end{equation}
As pointed out before, in linearized EC theory we have
\begin{equation}
    R^{\prime \,  {\rm lin}}_{\mu \nu \rho \sigma} = R^{\rm lin}_{\mu \nu \rho \sigma},
\end{equation}
but now we see that this condition is due to the fact that each term  in Eq. (\ref{eq:linearized_U4_Riemann}) is separately invariant, i.e., 
\begin{eqnarray}
\hat{R}^{\prime \, {\rm lin}}_{\mu \nu \rho \sigma} &=& \hat{R}^{\rm lin}_{\mu \nu \rho \sigma},  \label{eq:invariance_of_Riemann}
\\
\left(\partial_{\sigma} K_{\rho \nu \mu}\right)^{\prime} &=& \partial_{\sigma}K_{\rho \nu \mu}.
\end{eqnarray}

\subsection{Post-Newtonian expansion} \label{Sec:PN-expansion-EC}
In the following sections we  deal with the question of the generation of gravitational radiation by  spinning, weakly self-gravitating, slowly moving and weakly stressed sources, i.e., \emph{PN sources in EC theory}.
This means that we can exploit the  Blanchet-Damour approach (which has been summarized in Sec. \ref{sec:problem})  in order to solve approximately the mathematical problem (\ref{eq:math-problem-EC2}). Thus, we will first  compute the gravitational field in the interior domain $\mathfrak{D}_i$ of the source via the  PN method\footnote{A PN investigation similar to the one developed in Sec. \ref{sec:1PN} can be found in Refs. \cite{Castagnino1985,Castagnino1987}. However, our PN analysis  will consider both instantaneous and retarded potentials with the purpose of using, afterwards, the Blanchet-Damour formalism.}. In Sec. \ref{Sec:Preamble to the PN expansion} we give some preliminary remarks and discuss the PN order of the relevant quantities.  The 0PN limit is presented in Sec. \ref{sec:0PN}, while the details of the calculations regarding the 1PN inner metric and the conservation laws are reported in Secs. \ref{sec:1PN} and \ref{sec:PN-concervation-laws}, respectively. 

We recall that PN series are performed in terms of $c^{-1}$ by keeping $G$  fixed, whereas in the PM pattern the expansion parameter is $G$ and $c$ is fixed (see Ref. \cite{BlanchetDamour1988} for details). 
Furthermore, we stress that in our analysis of spinning  PN sources we do not suppose, unlike linearized EC theory, that torsion is weak. Indeed, strictly speaking, this hypothesis should be invoked in the exterior weak-field region $\mathfrak{D}_e$ where the MPM series is physically valid. However, the torsion field vanishes in $\mathfrak{D}_e$ since the  spin angular momentum tensor $\tau_{\mu \nu}^{\phantom{\mu \nu}\lambda}$ has compact support in $\Omega$ (see Eq. (\ref{eq:torsion&tau}); note that in the PN formalism the region $\Omega_{\rm ext}$  coincides with $\mathfrak{D}_e$).

\subsubsection{Preamble to the PN expansion}
\label{Sec:Preamble to the PN expansion}
At 1PN level, our  calculations can be  more appropriately carried out in terms of the metric tensor components $g_{\mu \nu}$, whereas at higher PN orders the use of the gravitational field amplitude $\mathfrak{h}^{\alpha\beta}$ is more convenient, likewise the GR case (see Refs. \cite{Blanchet-Damour1989,Damour-Iyer1991b,Blanchet1995,Blanchet2014}). Therefore, the starting point of our PN investigation is represented by Eq. (\ref{eq:Einstein-Cartan_equations})
worked out in harmonic gauge (\ref{eq:prob-EC2}). This is equivalent to consider Eq. (\ref{eq:proa-EC2}) in terms of $g_{\alpha \beta}$, see Eq. (\ref{eq:def_of_h_alphabeta2}). In this way, we get
\begin{align} \label{eq:Einstein-Cartan-with-gauge}
  & 2\hat{R}_{\alpha \beta}  =  \chi  \, \mathcal{G}_{\alpha \mu \beta \nu}\, T^{\mu \nu}  +  \dfrac{\chi^2}{2} \, \mathcal{G}_{\alpha \mu \beta \nu}\, \mathcal{S}^{\mu \nu},
\end{align}    
where 
\begin{align} \label{eq:hat-Ricci-in-harmonic-coordinates}
2\hat{R}_{\alpha \beta} &= -g^{\mu \nu} g_{\alpha \beta, \mu \nu} + g^{\mu \nu}g^{\rho \sigma} \Biggl[  g_{\alpha \mu,\rho} \left(2 g_{\beta [\nu,\sigma]}+ g_{\nu \sigma, \beta} \right)  
       \nonumber\\
       &+  g_{\beta \mu, \rho}g_{\nu \sigma, \alpha} - \dfrac{1}{2} g_{\mu \rho, \alpha} g_{\nu \sigma, \beta} \Biggr] 
     \\
 \mathcal{G}_{\alpha \mu \beta \nu} &\equiv g_{\alpha \mu} g_{\beta \nu} - \dfrac{1}{2} g_{\alpha \beta} g_{\mu \nu},
\end{align}
and $g_{\alpha \beta,\mu} \equiv \partial_\mu g_{\alpha \beta}$ (for a comparison, see Ref. \cite{Blanchet-Damour1989}). Moreover, crucial will be the inspection of the PN order of the various quantities occurring in Eq. (\ref{eq:Einstein-Cartan-with-gauge}). For this reason, let ${}^{(n)}{g}{_{\mu \nu}}$ denote the term in $g_{\mu \nu}$ of order $\left(v/c\right)^n$, whereas  ${}^{(n)}{T}{_{\mu \nu}}$ the contribution in ${T}{_{\mu \nu}}$ of order $\dfrac{M v^n}{d^3 c^{n-2}}$ (with $M$, $d$, and $v$  the mass, the typical linear dimension, and the internal velocity of the source, respectively). We also have for the torsion field $S_{\mu \nu}^{\phantom{\mu \nu}\lambda}$, the  spin angular momentum tensor $\tau_{\lambda}^{\phantom{\lambda}\mu \nu}$, and the  tensor $\mathcal{S}^{\mu \nu}$, respectively, 
\begin{align}
    {}^{(n)}S_{\mu \nu}^{\phantom{\mu \nu}\lambda} & \sim  \left(\dfrac{v}{c}\right)^n \, \dfrac{1}{d}, \nonumber\\
    {}^{(n)}\tau_{\lambda}^{\phantom{\lambda}\mu \nu} & \sim \dfrac{M}{d^2} \dfrac{v^n}{c^{n-2}},\nonumber\\  
    {}^{(n)}\mathcal{S}^{\mu \nu} & \sim  \dfrac{M^2}{d^4} \dfrac{v^{2n}}{c^{2n-4}}. 
\end{align}
By considering the following physical interpretation of the components of  $\tau_{\lambda}^{\phantom{\lambda}\mu \nu}$  \cite{Hehl1976_tensor}
\begin{align}
\label{eq: tau_physical}
        & \tau_{0}^{\phantom{\lambda}i 0}= \text{energy-dipole-moment density}, \nonumber\\
        & \tau_{0}^{\phantom{\lambda}i j}= (\text{energy-dipole-moment flux})/c, \nonumber\\
        & \tau_{i}^{\phantom{\lambda}j 0}= (\text{spin density})c, \nonumber\\
        & \tau_{i}^{\phantom{\lambda}jk}= \text{spin flux}, 
\end{align}
their PN order reads as 
\begin{align}
  & \tau_{0}^{\phantom{\lambda}i 0} = {\rm O}(c^2), \nonumber\\
        & \tau_{0}^{\phantom{\lambda}i j} = {\rm O}(c), \nonumber\\
        & \tau_{i}^{\phantom{\lambda}j 0} = {\rm O}(c), \nonumber\\
        & \tau_{i}^{\phantom{\lambda} j k} = {\rm O}(c^0), 
 \end{align}
and hence we get the following PN series:
\begin{align}
    \label{eq:tau_PN_expansion}
        & \tau_{0}^{\phantom{\lambda}i 0} ={}^{(0)}\tau_{0}^{\phantom{\lambda}i 0}+ {}^{(2)}\tau_{0}^{\phantom{\lambda}i 0} + \dots , \nonumber\\
        & \tau_{0}^{\phantom{\lambda}i j} ={}^{(1)}\tau_{0}^{\phantom{\lambda}i j}+ {}^{(3)}\tau_{0}^{\phantom{\lambda}i j} + \dots , \nonumber\\
    & \tau_{i}^{\phantom{\lambda} j 0} ={}^{(1)}\tau_{i}^{\phantom{\lambda}j 0}+ {}^{(3)}\tau_{i}^{\phantom{\lambda}j 0} + \dots , \nonumber\\
   & \tau_{i}^{\phantom{\lambda} j k} ={}^{(2)}\tau_{i}^{\phantom{\lambda}j k}+ {}^{(4)}\tau_{i}^{\phantom{\lambda}j k} + \dots .
\end{align}
Therefore, bearing in mind Eq. (\ref{Eq:S_alphabeta}), we obtain the following PN expansion for $\mathcal{S}^{\mu \nu}$:
\begin{align}
 \label{eq:S_expansion}
       & \mathcal{S}^{00}={}^{(0)}\mathcal{S}^{00} + {}^{(1)}\mathcal{S}^{00} +{}^{(2)}\mathcal{S}^{00} + \dots, \nonumber\\
       & \mathcal{S}^{0i}={}^{(0.5)}\mathcal{S}^{0i} + {}^{(1.5)}\mathcal{S}^{0i} +{}^{(2.5)}\mathcal{S}^{0i} + \dots, \nonumber\\  
            & \mathcal{S}^{ij}={}^{(0)}\mathcal{S}^{ij} + {}^{(1)}\mathcal{S}^{ij} +{}^{(2)}\mathcal{S}^{ij} + \dots.
\end{align}
This can be summarized by\footnote{ We are employing a standard notation in the literature (see e.g. \cite{Damour-Iyer1991b}) according to which
    \begin{align*}
    A^{\mu} & ={\rm O}(c^{-q},c^{-p}) \Leftrightarrow  A^{0}={\rm O}(c^{-q}), \, A^{i}={\rm O}(c^{-p}),
    \nonumber \\
       T^{\mu \nu} & ={\rm O}(c^{-q},c^{-p},c^{-r})  
       \nonumber \\
      & \Leftrightarrow T^{00}={\rm O}(c^{-q}), \, T^{0i}={\rm O}(c^{-p}), \, T^{ij}={\rm O}(c^{-r}) .
    \end{align*}}
\begin{equation} \label{eq:S_expansion_2}
  \mathcal{S}^{\alpha \beta} = {\rm O}\left(c^4,c^3,c^4\right). 
\end{equation}

Thanks to the above equation, the tensor $\Theta^{\alpha \beta}$, defined in Eq. (\ref{eq:tilde-T-alpha-beta}), has the same PN structure as the metric stress-energy tensor in GR, i.e., 
\begin{equation} \label{eq:PN-of-Theta}
  \Theta^{\alpha \beta} = {\rm O}\left(c^2,c,c^0\right).
\end{equation}
Therefore, we can expect that the PN analysis of Eq. (\ref{eq:hat-G-equals-tilde-T}) will yield for the metric $g_{\mu \nu}$ a PN expansion analogous to the GR case, as we will show in the next sections. 

\subsubsection{0PN expansion}
\label{sec:0PN}
The PN expansion (\ref{eq:S_expansion}) allows us to consider a 0PN limit for the metric tensor  represented by
\begin{subequations}\label{eq:0PN_metric}
\begin{align}
g_{00}&=-1+{}^{(2)}g_{00}+\rm{O}\left(c^{-4}\right),\label{eq:0PN_metric_a}
\\
g_{0i}&=\rm{O}\left(c^{-3}\right), \label{eq:0PN_metric_b}
\\
g_{ij}&=\delta_{ij}+\rm{O}\left(c^{-2} \right). \label{eq:0PN_metric_c}
\end{align}    
\end{subequations}
By exploiting  (\ref{eq:S_expansion_2}) and (\ref{eq:0PN_metric}), Eq.  (\ref{eq:Einstein-Cartan-with-gauge}) gives 
\begin{align}
 \label{eq:0PN_potential}
 & {}^{(2)}g_{00} = - 2 \phi,
 \\
&\bigtriangleup \phi=  \dfrac{\chi}{4}\, {}^{(0)} T^{00},
\label{eq:0PN_potential_2}
\end{align}
where  $\bigtriangleup \equiv \delta^{i j} \partial_i \partial_j$ is the  three-dimensional flat-space  Laplace operator. The solution of Eq. (\ref{eq:0PN_potential_2}), together with the boundary condition that the field $\phi$ vanishes at spatial infinity, is given by the \emph{Poisson integral}
\begin{equation} \label{eq:potential phi}
\begin{aligned}
    \phi(t,\boldsymbol{x})&=\Delta^{-1}\left(\frac{\chi}{4}\,{}^{(0)}T^{00}\right)\\
    &\equiv -\dfrac{G}{c^4} \int_{\mathbb{R}^3} \dfrac{{\rm d}^3\boldsymbol{y}}{\vert \boldsymbol{x}-\boldsymbol{y}\vert} {}^{(0)}T^{00}\left(t,\boldsymbol{y}\right),
\end{aligned}
\end{equation}
where $-c^2 \phi$ represents the (sign-reversed) gravitational potential.

\subsubsection{1PN expansion}
\label{sec:1PN}
The  1PN approximation of the inner metric can be obtained by solving iteratively Eq. (\ref{eq:Einstein-Cartan-with-gauge}) starting with the linearized result (\ref{eq:0PN_metric}). 

We first  consider the \emph{spatial components} (i.e., $\alpha=i,\beta=j$) of Eq. (\ref{eq:Einstein-Cartan-with-gauge}) and get the following results:
\begin{eqnarray} 
 2 \hat{R}_{ij} &=& - \Box \left({}^{(2)}g_{ij}\right) + {\rm O}(c^{-4}), \label{eq:hat-Rij-PN}
\\
 \chi \, \mathcal{G}_{i \mu j \nu} T^{\mu \nu}  &=&  \dfrac{\chi }{2} \delta_{ij}\left( T^{00}+T^{kk}\right)+ {\rm O}(c^{-4}), 
\\
 \dfrac{\chi^2}{2} \, \mathcal{G}_{i \mu j \nu} \mathcal{S}^{\mu \nu}  &=& \dfrac{\chi^2}{2} \left[\mathcal{S}_{ij}+\dfrac{1}{2}\delta_{ij}\left(\mathcal{S}^{00}-\mathcal{S}^{kk}\right)\right] + {\rm O}(c^{-6}) 
 \nonumber \\
 &=& {\rm O}(c^{-4}),
\label{eq:hat.Sij-PN}
\end{eqnarray}
and hence $g_{ij}$ satisfies the equation
\begin{equation}\label{eq:g_ij_eq_provvisoria}
    \Box g_{ij}= -\dfrac{\chi}{2}\delta_{ij}\left( T^{00}+T^{kk}\right)+ {\rm O}(c^{-4}).
\end{equation}
From the last equation, if we  analyze  only  ${\rm O}(c^{-2})$ contributions, we obtain the Poisson equation
\begin{equation}
    \bigtriangleup \left({}^{(2)}g_{ij}\right) = -\dfrac{\chi}{2} \delta_{ij} {}^{(0)}T_{00},
\end{equation}
which, by means of Eqs. (\ref{eq:0PN_potential})--(\ref{eq:potential phi}), leads to the same result as in Einstein theory, i.e.,
\begin{equation} \label{eq:1PN_gij_1}
    {}^{(2)}g_{ij} = \delta_{ij} {}^{(2)}g_{00} = -2 \phi \delta_{ij}.
\end{equation}
Therefore, we obtain the 1PN approximation (see Eq. (\ref{eq:0PN_metric_c}))
\begin{equation} \label{eq:g_ij_1PN_provvisoria}
    g_{ij}=\delta_{ij}-2\phi\delta_{ij}+{\rm O}(c^{-4}).
\end{equation}

The \emph{temporal component} (i.e., $\alpha=\beta=0$) of Eq.  (\ref{eq:Einstein-Cartan-with-gauge}) can be worked out, after some algebra, as follows:
\begin{eqnarray}
2 \hat{R}_{00} &=& - \Box \left({}^{(2)}g_{00}\right) -\Box\left({}^{(4)}g_{00}\right) 
    \nonumber \\
  &+& 2 \left({}^{(2)}g_{ij}{}^{(2)}g_{00,ij}\right)    -  \dfrac{1}{2}\bigtriangleup \left({}^{(2)}g_{00}\right)^2 
  \nonumber \\
  &+&  {\rm O}(c^{-6}), \label{eq:hat-R00-PN}
    \\
\chi \, \mathcal{G}_{0 \mu 0 \nu} T^{\mu \nu} &=& \chi \left[\dfrac{1}{2}\left(T^{00}+T^{kk}\right)-{}^{(2)}g_{00} T^{00} \right]
\nonumber \\
&+&  {\rm O}(c^{-6}), 
\\
 \dfrac{\chi^2}{2} \, \mathcal{G}_{0 \mu 0 \nu} \mathcal{S}^{\mu \nu} 
&=&   \dfrac{\chi^2}{2} \left[\dfrac{1}{2} \left(\mathcal{S}^{00}+\mathcal{S}^{kk}\right)\right] 
\nonumber \\
&+& {\rm O}(c^{-6}). \label{eq:hat-S00-PN}
\end{eqnarray}
Therefore, Eqs. (\ref{eq:hat-R00-PN})--(\ref{eq:hat-S00-PN}) give 
\begin{eqnarray} 
&&  - \Box \left({}^{(2)}g_{00}\right) -\Box\left({}^{(4)}g_{00}\right) + 2 
    \left({}^{(2)}g_{ij}{}^{(2)}g_{00,ij}\right)  
    \nonumber \\
&&  - \dfrac{1}{2}\bigtriangleup \left({}^{(2)}g_{00}\right)^2  = \chi \left[\dfrac{1}{2}\left(T^{00}+T^{kk}\right)-{}^{(2)}g_{00} T^{00} \right] 
\nonumber \\
&& +   \dfrac{\chi^2}{2} \left[\dfrac{1}{2} \left(\mathcal{S}^{00}+\mathcal{S}^{kk}\right)\right] + {\rm O}(c^{-6}). \label{eq:g00_eq_provvisoria}
\end{eqnarray}
By exploiting the result (cf. Eqs. (\ref{eq:0PN_potential}), (\ref{eq:0PN_potential_2})  and (\ref{eq:1PN_gij_1}))
\begin{align}
 & 2 \left({}^{(2)}g_{ij}{}^{(2)}g_{00,ij}\right) = 2 \left({}^{(2)}g_{00}{}^{(2)} g_{00,jj}\right) 
 \nonumber \\
 & = 2\;{}^{(2)}g_{00} \left( -\dfrac{\chi}{2}\, {}^{(0)}T^{00} \right),   
\end{align}
Eq. (\ref{eq:g00_eq_provvisoria}) can be further simplified, yielding 
\begin{align} \label{eq:g00_eq_simplified}
       &  - \Box \left({}^{(2)}g_{00}\right) -\Box\left({}^{(4)}g_{00}\right) - \dfrac{1}{2}\bigtriangleup \left({}^{(2)}g_{00}\right)^2 
       \nonumber \\
       & = \dfrac{\chi}{2} \left(T^{00}+T^{kk} \right) +   \dfrac{\chi^2}{4} \left(\mathcal{S}^{00}+\mathcal{S}^{kk}\right) + {\rm O}(c^{-6}).
\end{align}
By considering only ${\rm O}(c^{-4})$ terms, the above equation can be written as
\begin{align} \label{eq:psi_equation}
  \bigtriangleup \psi &= \phi_{,00} + \dfrac{\chi}{4} \left({}^{(2)}T^{00}+{}^{(2)}T^{kk} \right) 
  \nonumber \\
  &+   \dfrac{\chi^2}{8} \left({}^{(0)}\mathcal{S}^{00}+{}^{(0)}\mathcal{S}^{kk}\right),
\end{align}
where we have defined
\begin{equation} \label{eq:definition-of-psi}
    {}^{(4)}g_{00} + 2 \phi^2 \equiv -2 \psi.
\end{equation}
Owing to the asymptotic flatness property of the metric tensor $g_{\alpha\beta}$, the potential $\psi$ vanishes at spatial infinity, and the solution of Eq. (\ref{eq:psi_equation}) is 
\begin{align} \label{eq:psi-integral-1}
        \psi \left(t,\boldsymbol{x} \right) & =   -\dfrac{1}{4\pi}\int \dfrac{{\rm d}^3 \boldsymbol{y}}{\vert \boldsymbol{x}-\boldsymbol{y}\vert}\Biggl\{\dfrac{\partial^2 \phi}{c^2\partial t^2} +
  \dfrac{\chi}{4} \Bigl[{}^{(2)}{T}{^{00}}+ {}^{(2)}{T}{^{kk}}
  \nonumber \\
  & +\dfrac{\chi}{2} \left({}^{(0)}{\mathcal{S}}^{00} + {}^{(0)}{\mathcal{S}}^{kk} \right)\Bigr]\Biggr\}\left(t, \boldsymbol{y}\right).
\end{align}
From the above equations, we see that, at the 1PN order, $g_{00}$ reads as (see Eqs. (\ref{eq:0PN_metric_a}) and (\ref{eq:0PN_potential}))
\begin{equation} \label{g00-first-expression}
    g_{00}=-1-2\phi-2\left(\phi^2+\psi\right)+{\rm O}(c^{-6}).
\end{equation}

The 1PN approximation of the \emph{mixed components} (i.e., $\alpha=0, \beta=i$) of Eq.  (\ref{eq:Einstein-Cartan-with-gauge})  leads to  the following results:
\begin{eqnarray} 
 2 \hat{R}_{0i} &=& - \Box {}^{(3)} g_{0i} + {\rm O}(c^{-5}), 
\label{eq:hat-R0i-PN}
\\
  \chi  \, \mathcal{G}_{0 \mu i \nu} T^{\mu \nu} &=& -\chi \, T^{0i} + {\rm O}(c^{-5}), 
\\
 \label{eq:hat-S0i-PN}
    \dfrac{\chi^2}{2} \, \mathcal{G}_{0 \mu i \nu} \mathcal{S}^{\mu \nu} &=& - \dfrac{\chi^2}{2}   \mathcal{S}^{0i}+{\rm O}(c^{-7}).
\end{eqnarray}

In the above equations, if we disregard  ${\rm O}(c^{-5})$ terms, we obtain
\begin{equation} \label{eq:g-0i-no-fifth}
    \Box g_{0i}= \chi T^{0i} + {\rm O}(c^{-5}).
\end{equation}
Furthermore, if we ignore retardation effects in Eq. (\ref{eq:g-0i-no-fifth}), we obtain the Poisson equation
\begin{equation} \label{eq:Poisson-3-g_0i}
 \bigtriangleup \left({}^{(3)}g_{0i}\right) =\chi \, {}^{(1)}T^{0i},
\end{equation}
and we can write  $g_{0i}$ at 1PN order as (cf. Eq. (\ref{eq:0PN_metric_b}))
\begin{equation}\label{eq:g-0i-with-zeta-i}
    g_{0i}= \zeta_i + {\rm O}(c^{-5}),
\end{equation}
where, due to the asymptotic flatness of  $g_{\alpha\beta}$, the instantaneous potential $\zeta_i$ is subject to the boundary condition that it vanishes at spatial infinity  and hence, from Eq. (\ref{eq:Poisson-3-g_0i}), its expression  reads as
\begin{equation} \label{eq:integral-of-zeta_i-1}
\zeta_i (t, \boldsymbol{x}) = -\dfrac{4G}{c^4} \int \dfrac{{\rm d}^3 \boldsymbol{y}}{\vert \boldsymbol{x}- \boldsymbol{y}\vert} \, {}^{(1)}T^{0i} \left(t, \boldsymbol{y}\right).
\end{equation}

The 1PN results (\ref{eq:g_ij_1PN_provvisoria}), (\ref{g00-first-expression}) and (\ref{eq:g-0i-with-zeta-i}) depend on the instantaneous potentials $\phi(t, \boldsymbol{x})$, $\psi(t, \boldsymbol{x})$ and $\zeta_i(t, \boldsymbol{x})$, expressed by the Poisson integrals (\ref{eq:potential phi}),  (\ref{eq:psi-integral-1}) and  (\ref{eq:integral-of-zeta_i-1}), respectively.    On the other hand, the 1PN metric can also be written in terms of retarded potentials. In fact,  since $\phi={\rm O}(c^{-2})$ and $\psi = {\rm O}(c^{-4})$ (see Eqs. (\ref{eq:0PN_potential}) and (\ref{eq:definition-of-psi})), Eq. (\ref{g00-first-expression}) becomes
\begin{equation} \label{eq:g_00-1}
    g_{00}= -1 + \dfrac{2V}{c^2} - \dfrac{2V^2}{c^4} +{\rm O}(c^{-6}) =- {\rm e }^{-2V/c^2} +{\rm O}(c^{-6}),
\end{equation}
where we have defined the retarded potential $V \equiv -c^2 (\phi + \psi)$. Starting from Eqs. (\ref{eq:0PN_potential_2}) and (\ref{eq:psi_equation}), it is easy to show that the potential $V$ represents the  retarded solution of the following  equation:
\begin{equation} \label{eq:Box_V}
    \Box V = - 4 \pi G \sigma,
\end{equation}
and hence can be written as (cf. Eq. (\ref{eq:standard_retarded}))
\begin{align} \label{eq:definition-of-retarded-potential-V}
V(t,\boldsymbol{x})=G \int\dfrac{{\rm d}^3 \boldsymbol{y}}{\vert \boldsymbol{x}- \boldsymbol{y}\vert} \sigma \left(t- \vert \boldsymbol{x}- \boldsymbol{y}\vert/c, \boldsymbol{y}\right),  
\end{align}
where we have defined
\begin{equation}  \label{eq:sigma_tilde}
    \sigma \equiv \dfrac{\Theta^{00}+\Theta^{kk}}{c^2} =\dfrac{T^{00}+T^{kk}}{c^2} + \left(8 \pi G\right) \dfrac{\mathcal{S}^{00}+\mathcal{S}^{kk}}{c^6}.
\end{equation}

Using Eqs. (\ref{eq:g_00-1}) and (\ref{eq:Box_V}), Eq. (\ref{eq:g00_eq_simplified}) reads as
\begin{align}
    \Box \log (-g_{00})&= \dfrac{\chi}{2} \left(T^{00}+T^{kk} \right) 
    \nonumber \\
    & +   \dfrac{\chi^2}{4} \left(\mathcal{S}^{00}+\mathcal{S}^{kk}\right) + {\rm O}(c^{-6}).
\end{align}

By introducing terms beyond 1PN order,   Eq. (\ref{eq:g_ij_1PN_provvisoria}) becomes
\begin{equation} \label{g-ij-first-expression}
    g_{ij}= \delta_{ij} \left(1+\dfrac{2V}{c^2}\right) + {\rm O}(c^{-4}),
\end{equation}
which means that $g_{ij}$ satisfies the equation (see Eqs. (\ref{eq:Box_V})--(\ref{eq:sigma_tilde}))
\begin{align} \label{eq:Box-g-ij-beyond1PN}
    \Box g_{ij} &= -\dfrac{\chi}{2}\delta_{ij}\left( T^{00}+T^{kk}\right) 
    \nonumber \\
    &- \dfrac{\chi^2}{4} \delta_{ij} \left(\mathcal{S}^{00}+\mathcal{S}^{kk}\right)+{\rm O}(c^{-4}).
\end{align}

Furthermore, thanks to Eqs. (\ref{eq:hat-R0i-PN})--(\ref{eq:hat-S0i-PN}),  Eq. (\ref{eq:g-0i-no-fifth}) can be replaced by  
\begin{equation}\label{eq:g0i_eq_2}
    \Box g_{0i}= \chi\, T^{0i} +\dfrac{\chi^2}{2}\,\mathcal{S}^{0i}+ {\rm O}(c^{-5}). 
\end{equation}
Note that in the above equation terms containing torsion contribution are beyond 1PN level (cf. Eq. (\ref{eq:S_expansion})). However, we have decided to include them in Eq. (\ref{eq:g0i_eq_2}) since this choice will be convenient for the subsequent calculations. Therefore, Eq. (\ref{eq:g-0i-with-zeta-i}) can be equivalently written  as
\begin{equation} \label{eq:g-01-at-1PN-1}
    g_{0i}= - \dfrac{4}{c^3} V_i + {\rm O}(c^{-5}),
\end{equation}
where the  potential $V_i$ represents the retarded solution of the (flat-space) wave  equation
\begin{equation}\label{eq:Box_V_i}
    \Box V_{i} = - 4 \pi G \sigma_i,
\end{equation}
and hence it can be written as
\begin{equation}
    \begin{split} \label{eq:potential-V_i-def}
     V_i\left(t,\boldsymbol{x} \right) =  G \int \dfrac{{\rm d}^3 \boldsymbol{y}}{\vert \boldsymbol{x}- \boldsymbol{y}\vert} \sigma_i \left(t- \vert \boldsymbol{x}- \boldsymbol{y}\vert/c, \boldsymbol{y}\right),
    \end{split}
\end{equation}
where 
\begin{equation} \label{eq:sigma_i}
   \sigma_i \equiv  \dfrac{\Theta^{0i}}{c} = \dfrac{T^{0i}}{c}+\left(8 \pi G\right) \dfrac{\mathcal{S}^{0i}}{c^5}.
\end{equation}

\subsubsection{PN conservation laws}\label{sec:PN-concervation-laws}
From the conservation law (\ref{eq:nabla_tilde_T}) we obtain
\begin{align} 
    & \partial_t \sigma + \partial_i \sigma_i = \dfrac{1}{c^2}\left(\partial_t \sigma_{jj} -\sigma  \partial_t V\right)+{\rm O}(c^{-4}),
    \label{eq:conservation_law} 
    \\
 & \partial_t \sigma_i + \partial_j \sigma_{ij} = \sigma \partial_i  V+{\rm O}(c^{-2}),  
   \label{eq:conservation_law_number-2} 
\end{align}
where,  for our purposes, we only need Newtonian  accuracy in Eq. (\ref{eq:conservation_law_number-2}), and we have defined
\begin{equation}\label{eq:tilde-sigma-ij}
    \sigma_{ij} = \Theta^{ij} = T^{ij}+ \left(8 \pi G\right)\dfrac{\mathcal{S}^{ij}}{c^4}.
\end{equation}
In Eqs. (\ref{eq:conservation_law}) and (\ref{eq:conservation_law_number-2}) we can replace 
the retarded potential $V$ with the instantaneous potential $U$, which is defined by the expansion of Eq. (\ref{eq:definition-of-retarded-potential-V}) for small retardation effects. This calculation gives 
\begin{equation} \label{eq:V-in-terms-U-and-X}
    V = U + \dfrac{1}{2c^2} \partial^2_t X + {\rm O}\left(c^{-3}\right),
\end{equation}
where 
\begin{equation}
    \begin{split}
    & U\left(t,\boldsymbol{x} \right) = G \int \dfrac{{\rm d}^3 \boldsymbol{y}}{\vert \boldsymbol{x}- \boldsymbol{y}\vert} \sigma \left(t, \boldsymbol{y}\right),
    \end{split}
\end{equation}
satisfies the Poisson equation 
\begin{equation}
   \bigtriangleup U = -4 \pi G \sigma,  
\end{equation}
with  the boundary condition, following from the asymptotic flatness of the metric tensor $g_{\alpha\beta}$, that it approaches zero at spatial infinity, whereas
\begin{equation} \label{eq:X-field-torsion}
    X(t,\boldsymbol{x})= G \int {\rm d}^3 \boldsymbol{y}\, \vert \boldsymbol{x}- \boldsymbol{y}\vert \, \sigma \left(t, \boldsymbol{y}\right),
\end{equation}
is known in the literature as \emph{super-potential} (i.e., a potential sourced by another potential, see Refs. \cite{Blanchet-Faye-Whiting2014,Poisson-Will2014} for further details). 

Similarly, the retarded potential $V_i$ can be expanded as (see Eq. (\ref{eq:potential-V_i-def}))
\begin{equation} \label{eq:PN-expansion-of-V_i} 
    V_i = U_i + {\rm O}\left(c^{-2}\right),
\end{equation}
where the instantaneous potential $U_i$ is the solution of the Poisson equation 
\begin{equation}
   \bigtriangleup U_i = -4 \pi G \sigma_i,  
\end{equation}
and,  being subject to the boundary condition that it vanishes at spatial infinity (as demanded by the asymptotic flatness of   $g_{\alpha\beta}$), can be written as
\begin{equation}
    \begin{split}
    & U_i\left(t,\boldsymbol{x} \right) = G \int \dfrac{{\rm d}^3 \boldsymbol{y}}{\vert \boldsymbol{x}- \boldsymbol{y}\vert} \sigma_i \left(t, \boldsymbol{y}\right).
    \end{split}
\end{equation}
Note that $U_i = -(c^3/4) \zeta_i + {\rm O}(c^{-2})$ (cf. Eq. (\ref{eq:integral-of-zeta_i-1})). 

If we ignore terms $O(c^{-2})$ in Eq. (\ref{eq:conservation_law}) and take into account Eqs. (\ref{eq:0PN_potential_2}) and (\ref{eq:Poisson-3-g_0i}), we obtain
\begin{equation}
    4 \partial_0 \phi + \partial_i \zeta_i=0,
\end{equation}
which in turn is equivalent to the gauge condition (\ref{eq:prob-EC2}). 

\subsection{Multipole analysis of the inner metric}\label{Sec:Multipole_analysis_inner_torsion}
Let us recall from Sec. \ref{sec:1PN} that the inner metric at 1PN order reads as (cf. Eqs. (\ref{eq:g_00-1}),  (\ref{g-ij-first-expression}) and (\ref{eq:g-01-at-1PN-1})) 
\begin{align} 
g^{\rm in}_{00} &= - {\rm e}^{-2V^{\rm in}/c^2} + {\rm O}(c^{-6}), 
\nonumber \\
g^{\rm in}_{0i} &= -\dfrac{4}{c^3} V^{\rm in}_i + {\rm O}(c^{-5}), 
\nonumber \\
g^{\rm in}_{ij} &= \delta_{ij} \left(1 + \dfrac{2}{c^2}V^{\rm in}\right)+ {\rm O}(c^{-4}), 
\label{eq:1PN_metric_torsion}
\end{align}
where hereafter we indicate explicitly with the superscript \qm{in} that we are investigating  the inner field. Since  $\sigma$ and $\sigma_i$ have compact support, we can perform the multipole expansion of the retarded solutions of the wave equations (\ref{eq:Box_V}) and (\ref{eq:Box_V_i})   in the region outside their sources, i.e., in the overlapping domain $\mathfrak{D}_o$. Exploiting the results of Appendix B of Ref. \cite{Blanchet-Damour1989} and the assumption of slow internal motions of the matter distribution,  we obtain, in the  Blanchet-Damour multi-index notation (see footnote \ref{footnote-multindex}), 
\begin{align}
 V^{\rm in} &= G \sum_{l=0}^{+ \infty} \dfrac{(-1)^l}{l!} \partial_L \left(\dfrac{\lambda_L(u)}{r}\right)+ {\rm O}(c^{-4}), 
 \label{eq:V-in-EC-1}
 \\
V^{\rm in}_i  & = -G \sum_{l=1}^{+ \infty} \dfrac{(-1)^l}{l!} \Biggl\{ \partial_{L-1} \left(\dfrac{K_{iL-1}(u)}{r}\right) 
\nonumber \\
& + \dfrac{l}{l+1} \epsilon_{iab}\partial_{aL-1} \left(\dfrac{J_{bL-1}(u)}{r}\right)\Biggr \}
\nonumber \\
&+  G \sum_{l=1}^{+ \infty}  \dfrac{(-1)^l}{l!} \dfrac{2l-1}{2l+1} \partial_{iL-1} \left(\dfrac{\mu_{L-1}(u)}{r}\right)+ {\rm O}(c^{-2}),
\label{eq:V-i-in-EC-1}
\end{align}
where $u\equiv t-r/c$ and ($ \widecheck{y}_L = y_{<L>}$ denotes the STF projection of $y_L$)
\begin{align}
\lambda_L(u) &\equiv \int {\rm d}^3 \boldsymbol{y} \, y_{<L>}\, \sigma(\boldsymbol{y},u) 
\nonumber \\
& + \dfrac{1}{2(2l+3)}\dfrac{1}{c^2} \dfrac{{\rm d}^2}{{\rm d}u^2} \int {\rm d}^3 \boldsymbol{y} \, y_{<L>} \,  \boldsymbol{y}^2 \sigma(\boldsymbol{y},u), \label{eq:Lambda_L-torsion}
\\
 K_L(u) & \equiv l \int {\rm d}^3 \boldsymbol{y} \, \widecheck{y}_{<L-1} \sigma_{i_l>} (\boldsymbol{y},u), 
 \\
 J_L(u) & \equiv \int {\rm d}^3 \boldsymbol{y} \, \epsilon_{ab<i_l} \widecheck{y}_{L-1>a} \sigma_b(\boldsymbol{y},u),\label{eq:J_L_torsion} 
 \\
\mu_L(u) & \equiv \int {\rm d}^3 \boldsymbol{y} \, y_{<iL>} \sigma_i(\boldsymbol{y},u).
\label{eq:Mu_L-torsion}
\end{align}
Note that, according to the recipe  of Sec. \ref{sec:resolution}, Eqs. (\ref{eq:V-in-EC-1}) and (\ref{eq:V-i-in-EC-1}) should be denoted with $\mathcal{M}(V^{\rm in})$ and $\mathcal{M}(V^{\rm in}_i)$, respectively. Despite that, hereafter we  do not follow this convention 
in order to ease the notation, since the operation of taking the multipole expansion will always be clear from the context. 

At this stage, we perform a coordinate transformation from the source-covering coordinate system $x^\alpha$ to the new coordinate system $x^{\prime \alpha}$, \emph{a priori} valid only in  $\mathfrak{D}_o$. Therefore, we have \cite{Blanchet-Damour1989}
\begin{align}\label{eq:coord-transf-torsion-x-x-prime}
x^{\prime 0} &    = x^0 - \dfrac{4G}{c^3} \sum_{l=0}^{+ \infty} \dfrac{(-1)^l}{(l+1)!} \dfrac{2l+1}{2l+3} \partial_L \left(\dfrac{\mu_L(u)}{r}\right), 
\nonumber \\
x^{\prime i } &   = x^i.
\end{align}
Such a transformation does not spoil the gauge condition (\ref{eq:prob-EC2}) and hence it guarantees that the new inner metric $g_{\mu \nu}^{\prime \, {\rm in}}(x^{\prime})$ has the same functional form of $g_{\mu \nu}^{{\rm in}}(x)$ (see Eq. (\ref{eq:1PN_metric_torsion})), i.e., 
\begin{align} 
g^{\prime \, {\rm in}}_{00} &= - {\rm e}^{-2V^{\prime \, {\rm in}}/c^2} + {\rm O}(c^{-6}), 
\nonumber \\
g^{\prime \, {\rm in}}_{0i} &= -\dfrac{4}{c^3} V^{\prime \, {\rm in}}_i + {\rm O}(c^{-5}), 
\nonumber \\
g^{\prime \, {\rm in}}_{ij} &= \delta_{ij} \left(1 + \dfrac{2}{c^2}V^{\prime \, {\rm in}}\right)+ {\rm O}(c^{-4}).
\label{eq:1PN-primed_metric_torsion}
\end{align}
The only difference with $g_{\mu \nu}^{{\rm in}}(x)$ is that the metric (\ref{eq:1PN-primed_metric_torsion}) is parametrized in terms of new scalar and vector potentials $V^{\prime \, {\rm in}}$ and $V_i^{\prime \, {\rm in}}$ depending, unlike Eqs. (\ref{eq:V-in-EC-1}) and (\ref{eq:V-i-in-EC-1}),  on only two families
 of STF tensors, i.e., the source multipole moments $I_L$ and $J_L$. Indeed, we have (suppressing for simplicity the primes on the new coordinates)
 \begin{align}
    V^{\prime \, {\rm in}} &= G \sum_{l=0}^{+ \infty} \dfrac{(-1)^l}{l!} \partial_L \left(\dfrac{I_L(u)}{r}\right)+{\rm O}(c^{-4}),
    \label{V-in-prime-torsion}
\\
   V^{\prime \, {\rm in}}_i  & = -G \sum_{l=1}^{+ \infty} \dfrac{(-1)^l}{l!} \Biggl\{ \partial_{L-1} \left(\dfrac{\dot{I}_{iL-1}(u)}{r}\right) 
   \nonumber \\
   & + \dfrac{l}{l+1} \epsilon_{iab}\partial_{aL-1} \left(\dfrac{J_{bL-1}(u)}{r}\right)\Biggr \} + {\rm O}(c^{-2}),
   \label{V_i-in-prime-torsion}
\end{align}
where $\dot{I}_L(u) \equiv {\rm d}I_L(u)/{\rm d}u$. In Eq. (\ref{V-in-prime-torsion}), $I_L$ is given by 
\begin{equation}
    I_L(u) \equiv   \lambda_L(u) - \dfrac{4(2l+1)}{(l+1)(2l+3)}\dfrac{1}{c^2} \dot{\mu}_L(u),
\end{equation}
and hence (cf. Eqs. (\ref{eq:Lambda_L-torsion}) and (\ref{eq:Mu_L-torsion}))
\begin{align} 
 \label{eq:I_L_torsion}
I_L(u) & = \int {\rm d}^3 \boldsymbol{y} \, y_{<L>} \sigma(\boldsymbol{y},u) 
\nonumber\\
& + \dfrac{1}{2(2l+3)}\dfrac{1}{c^2} \dfrac{{\rm d}^2}{{\rm d}u^2} \int {\rm d}^3 \boldsymbol{y} \, y_{<L>} \,  \boldsymbol{y}^2 \sigma(\boldsymbol{y},u) 
\nonumber \\
& - \dfrac{4(2l+1)}{(l+1)(2l+3)}\dfrac{1}{c^2} \dfrac{{\rm d}}{{\rm d}u} \int {\rm d}^3 \boldsymbol{y} \, y_{<iL>} \sigma_i (\boldsymbol{y},u),
\end{align}
while, in Eq. (\ref{V_i-in-prime-torsion}), $J_L$ is given by Eq. (\ref{eq:J_L_torsion}) and we have exploited  the conservation law (\ref{eq:conservation_law}) up to ${\rm O}(c^{-2})$ terms to write
\begin{equation} \label{eq:K-L-multipole-torsion}
    K_L(u) = \dot{I}_{L}(u) + {\rm O}(c^{-2}). 
\end{equation}

It is worth noting that in this section the source multipole moments $I_L$ and $J_L$   parametrize the multipole decomposition of the inner field, as an inspection of Eqs. (\ref{eq:1PN-primed_metric_torsion})--(\ref{V_i-in-prime-torsion}) reveals, whereas in Sec. \ref{sec:resolution} they  characterize  the  exterior field (cf. Eq. (\ref{eq:source_multipole_moments})). However, this difference with respect to the GR procedure will not alter the outcome of our analysis, as we will show. 

Until now we have expressed, within $\mathfrak{D}_o$,  the 1PN inner metric $g_{\mu \nu}^{\prime \, {\rm in}}$ in terms of the source multipole moments $I_L$ and $J_L$, given in Eqs. (\ref{eq:I_L_torsion}) and (\ref{eq:J_L_torsion}), respectively. In the next section we will study  the external gravitational field with the purpose of relating, via the matching procedure,  $I_L$ and $J_L$ to the radiative moments $U_L$ and $V_L$ appearing in the asymptotic metric (cf. Sec. \ref{sec:resolution}). We will see that a crucial role in this framework will be fulfilled by the the canonical moments $M_L$ and $S_L$.

\subsection{External metric}\label{Sec:External_Metric_and_Matching}
Since the tensor field $\Theta^{\alpha \beta}$ has compact support in $\Omega$, the mathematical problem (\ref{eq:math-problem-EC2}) assumes in the outer  region $\mathfrak{D}_e$ the same form as in Einstein theory  and hence can be addressed by resorting to the MPM algorithm devised in GR (see Sec. \ref{sec:resolution}). The output of this procedure is the canonical external metric, which is parametrized  in terms of the canonical multipole moments $M_L$ and $S_L$. In order to perform the matching, which allows  to relate $M_L$ and $S_L$ to the source moments $I_L$ and $J_L$, expressed by Eqs. (\ref{eq:I_L_torsion}) and (\ref{eq:J_L_torsion}), respectively, it is essential to first  consider  the PN expansion, in $\mathfrak{D}_o$, of the canonical external metric. However, we stress once again that, at this level, we only need to exploit an \qm{order by order} matching procedure, so a less general pattern than the one outlined in Sec. \ref{sec:resolution} \cite{Blanchet1995,Blanchet1998,BlanchetDamour1988}.

When the external metric is expanded in a PN fashion in the matching region $\mathfrak{D}_o$,
each term $\mathfrak{h}^{\alpha \beta}_{(n)}$ (cf. Eq. (\ref{PM_expansion1})) has, according to the results of Ref. \cite{Blanchet-Damour1986}, the following structure: 
\begin{equation}
\mathfrak{h}^{\alpha \beta}_{(n)}= {\rm O}\left(c^{-2n},c^{-2n-1},c^{-2n} \right).
\end{equation}
This means that in order to perform the matching to the inner field we need to go one step beyond the linearized result in the MPM algorithm, i.e., we need to compute  $\mathfrak{h}^{\alpha \beta}_{(1)}$ and $\mathfrak{h}^{\alpha \beta}_{(2)}$ (see Eq. (\ref{PM_expansion1})). In this way,  as shown in details in Ref. \cite{Blanchet-Damour1989}, we end up with the following 1PN (re-)expansion for the canonical external metric $g_{\alpha \beta}^{\rm ext}$ valid in $\mathfrak{D}_o$:
\begin{align} 
g^{\rm ext}_{00} &= - {\rm e}^{-2V^{\rm ext}/c^2}+{\rm O}(c^{-6}), 
\nonumber \\
g^{\rm ext}_{0i} &= -\dfrac{4}{c^3} V^{\rm ext}_i + {\rm O}(c^{-5}), 
\nonumber\\
g^{\rm ext}_{ij} &= \delta_{ij} \left(1 + \dfrac{2}{c^2}V^{\rm ext}\right)+ {\rm O}(c^{-4}), 
\label{PN_re-expansion_ext_field}
\end{align}
where the external potentials $V^{\rm ext}$ and  $V_i^{\rm ext}$, kept in PM form, are given in terms of the canonical moments $M_L$ and $S_L$ by
\begin{align} 
V^{\rm ext} &= G \sum_{l=0}^{+ \infty} \dfrac{(-1)^l}{l!} \partial_L \left(\dfrac{M_L(u)}{r}\right), 
\nonumber \\
V^{\rm ext}_i &= -G \sum_{l=1}^{+ \infty} \dfrac{(-1)^l}{l!} \Biggl\{ \partial_{L-1} \left(\dfrac{\dot{M}_{iL-1}(u)}{r}\right) 
\nonumber \\
& + \dfrac{l}{l+1} \epsilon_{iab}\partial_{aL-1} \left(\dfrac{S_{bL-1}(u)}{r}\right)\Biggr \}.
\label{eq:external_metric_torsion}
\end{align}

It should be stressed that in Eq. (\ref{PN_re-expansion_ext_field})  no tails effects appear. This result agrees with the theorem stated in Sec. V of Ref. \cite{BlanchetDamour1988} according to which such hereditary terms, denoted as $\mathfrak{t}^{\alpha \beta}_{(n)}$ (with $n\geq2$, the index $n$ occurring in $\mathfrak{t}^{\alpha \beta}_{(n)}$ being the same as the one labelling $\mathfrak{h}^{\alpha \beta}_{(n)}$ in Eq. (\ref{PM_expansion1})), quickly become negligible in the near zone $\mathfrak{D}_i$ when the index $n$ increases.  In particular, when $n=2$, the tail  $\mathfrak{t}^{\alpha \beta}_{(2)}$ occurring in $\mathfrak{h}^{\alpha \beta}_{(2)}$ is such that, in $\mathfrak{D}_i$, 
\begin{equation}
\mathfrak{t}^{\alpha \beta}_{(2)} = {\rm O}\left(\dfrac{\log c}{c^8}\right),
\end{equation} 
which is far beyond the remainder of Eq. (\ref{PN_re-expansion_ext_field}).

\subsection{Matching of the internal and external fields} \label{Sec:Matching-procedure}
We now have all the elements to  apply the matching procedure. It consists in requiring that the internal field $g_{\mu \nu}^{\prime\,{\rm in}}$, constructed in Eq. (\ref{eq:1PN-primed_metric_torsion}), and the near-zone-expanded external metric $g_{\mu \nu}^{\rm ext}$, given in Eq. (\ref{PN_re-expansion_ext_field}), should be isometric in their common domain of validity, i.e., the overlapping region $\mathfrak{D}_o$. Let $x^{\prime \mu}_{\rm in}$ be the harmonic coordinates \emph{a priori} valid only in the exterior part $\mathfrak{D}_o$ of the inner domain $\mathfrak{D}_i$ and related to the source-covering coordinates $x^{\mu}_{\rm in}$ used in  $\mathfrak{D}_i$ by the transformation  (\ref{eq:coord-transf-torsion-x-x-prime}). Upon denoting with $x^{\mu}_{\rm ext}$ the coordinates employed in the outer region $\mathfrak{D}_e$, the matching pattern requires the existence of a (PN-expanded) coordinate transformation, described through the map  
\begin{equation} \label{eq:coordinate_transf_2}
  \mathscr{F}:  x^{\prime \mu}_{\rm in} \rightarrow x^{\mu}_{\rm ext} = x^{\prime \mu}_{\rm in} - \varphi^{\prime \mu}(x^{\prime}_{\rm in}), 
\end{equation}
such that the pullback $\mathscr{F}^\star g^{\rm ext}$ of $g^{\rm ext}$ by $\mathscr{F}$ gives \cite{Nakahara2003,Wald}
\begin{equation}
    \left(\mathscr{F}^{\star} g^{\rm ext}\right)_{\mu \nu} = g^{\prime \, {\rm in}}_{\mu \nu}.
\end{equation}
The above equation,  worked out explicitly, leads to the \emph{exact} relation
\begin{equation} \label{eq:exact-rel-isometry}
    \dfrac{\partial x^\alpha_{\rm ext}}{\partial x^{\prime \mu}_{\rm in}} \dfrac{\partial x^\beta_{\rm ext}}{\partial x^{\prime \nu}_{\rm in}}\,g^{\rm ext}_{\alpha \beta} \left(x_{\rm ext}\right)= g^{\prime \, {\rm in}}_{\mu \nu} \left(x^\prime_{\rm in}\right).
\end{equation}

The vector $\varphi^{\prime \mu}(x^{\prime}_{\rm in})$ occurring in Eq. (\ref{eq:coordinate_transf_2}) admits a multipolar and PN expansion appropriate in $\mathfrak{D}_o$ and  is supposed to be of order (see e.g. Refs. \cite{BlanchetDamour1988,Blanchet1995,Damour-Iyer1991b}) 
\begin{align} 
    \varphi^{\prime \mu} &= {\rm O}(c^{-3},c^{-4}). \label{eq:vector_varphi}
   \end{align}
 
Starting from  Eq. (\ref{eq:exact-rel-isometry}), jointly with Eqs. (\ref{eq:coordinate_transf_2}) and  (\ref{eq:vector_varphi}), it is easy to see that the effects of the coordinate transformation (\ref{eq:coordinate_transf_2}) reduce to those of a standard linearized gauge transformation (see Sec. \ref{sec:Linearized-EC}) up to ${\rm O}(c^{-6},c^{-7},c^{-6})$ contributions, i.e.,
\begin{equation}\label{eq:effect_of_transformation_on_g}
    g_{\mu \nu}^{\rm ext}(x^\prime_{\rm in})-\partial_\mu \varphi^{\prime}_{ \nu} - \partial_\nu \varphi^{\prime}_{ \mu} + {\rm O}(c^{-6},c^{-7},c^{-6})=g_{ \mu \nu}^{\prime\,{\rm in}}(x^\prime_{\rm in}), 
\end{equation}
where we have expressed both sides of the equation in terms of the inner coordinates $x^\prime_{\rm in}$ and the error terms turn out to be even better than what  needed \emph{a priori} (i.e., ${\rm O}(c^{-6},c^{-7},c^{-6})$ instead of ${\rm O}(c^{-6},c^{-5},c^{-4})$, see Eqs. (\ref{eq:1PN-primed_metric_torsion}) and (\ref{PN_re-expansion_ext_field})). It should be stressed that the result expressed by Eq. (\ref{eq:effect_of_transformation_on_g}) could be expected \emph{a priori} since  the inner and outer fields  take the same functional up to  ${\rm O}(c^{-6},c^{-5},c^{-4})$ order, see Eqs. (\ref{eq:1PN-primed_metric_torsion}) and  (\ref{PN_re-expansion_ext_field}).

As a consequence  of Eq. (\ref{eq:effect_of_transformation_on_g}), we can use the invariance of the linearized Riemann tensor $\hat{R}^{\rm lin}_{\mu \nu \rho \sigma}$, which has been demonstrated in Eq. (\ref{eq:invariance_of_Riemann}), to write the following  identity for the  components $\hat{R}^{\rm lin}_{0i0j}$:  
\begin{align} \label{eq:effect of Riemann invariance}
       & \partial_{i}\partial_{j} \left(g^{\rm ext}_{00}-g^{\prime \,{\rm in}}_{00}\right)-\partial_0 \partial_i \left(g^{\rm ext}_{0j}-g^{\prime\, {\rm in}}_{0j}\right) 
       \nonumber \\
        & - \partial_0 \partial_j \left(g^{\rm ext}_{0i}-g^{\prime\, {\rm in}}_{0i}\right)+\partial_0 \partial_0 \left(g^{\rm ext}_{ij}-g^{\prime\, {\rm in}}_{ij}\right) =0.
\end{align}
The above equation permits obtaining the sought-after relation between the canonical moments $M_L$ and $S_L$  occurring in Eq. (\ref{eq:external_metric_torsion}) and the source moments $I_L$ and $J_L$ (cf. Eqs. (\ref{eq:I_L_torsion}) and (\ref{eq:J_L_torsion})) parametrizing the inner metric, i.e.,
\begin{align}\label{eq:canonical-source-moments-EC-theory}
        M_L(u) & =I_L(u) + {\rm O}(c^{-4}),
        \nonumber \\
         S_L(u) & =J_L(u) + {\rm O}(c^{-2}).
\end{align}
We point out that the remainder occurring in Eq. (\ref{eq:canonical-source-moments-EC-theory}) agrees with the one appearing in Eq. (3.25) of Ref. \cite{Blanchet-Damour1989}.

In the context of Einstein theory,  the analysis of the structure of $g_{\mu \nu}^{\rm ext}$ in the asymptotic wave zone $\mathfrak{D}_w$ has been performed in Ref. \cite{Blanchet1986} and has revealed that it is always possible to define a set of  radiative coordinates $X^{\mu} = (c T, \boldsymbol{X})$ where the external metric  admits an asymptotic expansion in powers of $R^{-1}$ at future null infinity (with $R = \vert \boldsymbol{X}\vert \equiv \left(\delta_{ij} X^i X^j \right)^{1/2} $). The same conclusions can be trivially drawn also  for our problem, since  in $\mathfrak{D}_e$, where both the  spin angular momentum tensor (\ref{eq:canonical_spin_tensor}) and the torsion tensor vanish, EC theory is formally analogous to  GR.  Therefore, following the calculations of Ref.  \cite{Blanchet-Damour1989}, we can easily obtain the relations between the  radiative moments and the canonical moments, i.e., (with $l \geq 2$)
\begin{align} \label{eq:radiative-canonical_1PN}
       U_L(u) &= \overset{(l)}{M}_L(u)+ {\rm O}(c^{-3}),
       \nonumber\\
       V_L(u) &= \overset{(l)}{S}_L(u) + {\rm O}(c^{-3}),
\end{align}
where superscript $(l)$ denotes the $l$-th time derivative of the moments with respect to the $u$ variable and the ${\rm O}(c^{-3})$ stands for the  tails effects. 

Equivalently, in terms of the source moments, from Eq. (\ref{eq:canonical-source-moments-EC-theory}) we have (with $l \geq 2$)
\begin{align} \label{eq:radiative-source_1PN}
       U_L(u) &= \overset{(l)}{I}_L(u)+ {\rm O}(c^{-3}),
       \nonumber \\
       V_L(u) &= \overset{(l)}{J}_L(u) + {\rm O}(c^{-2}).
\end{align}
Equations (\ref{eq:radiative-canonical_1PN}) and (\ref{eq:radiative-source_1PN}) represent the final result of the matching procedure and solve the GWs generation problem in EC model at 1 PN level. 

\subsection{The lowest-order source multipole moments} \label{Sec:The lowest-order source multipole moments}
The lowest-order source multipole moments can be read off from Eqs. 
(\ref{eq:J_L_torsion}) and (\ref{eq:I_L_torsion}) (see also Eq. (\ref{eq:canonical-source-moments-EC-theory})). By exploiting Eqs. (\ref{eq:conservation_law}), (\ref{eq:conservation_law_number-2}) and (\ref{eq:V-in-terms-U-and-X}) along with the \emph{Gauss theorem}, which allows to discard  integrals containing a total divergence, a lengthy calculation yields the following expressions:   
\begin{align}
    I\left(t\right) &= \int {\rm d}^3 \boldsymbol{y} \left[ \sigma + \dfrac{1}{c^2}\left(\dfrac{1}{2} \sigma U^{\rm in} -\sigma_{jj}\right)\right] + {\rm O}(c^{-4}), \label{eq:I-expression-torsion}
\\
    I_i \left(t\right)&= \int {\rm d}^3 \boldsymbol{y}\,y_i \left[ \sigma + \dfrac{1}{c^2}\left(\dfrac{1}{2} \sigma U^{\rm in} -\sigma_{jj}\right)\right] + {\rm O}(c^{-4}), \label{eq:I-i-expression-torsion}
\\
    J_i \left(t\right)&= \int {\rm d}^3 \boldsymbol{y}\,  \epsilon_{ijk} y_j \sigma_k + {\rm O}(c^{-2}). \label{eq:J-i-expression-torsion}
\end{align}
Moreover, we can introduce the following quantity: 
\begin{equation}\label{eq:P-i-expression-torsion}
    \mathscr{P}_i \left(t\right) \equiv \dfrac{{\rm d}}{{\rm d}t}I_i \left(t\right)= \int {\rm d}^3 \boldsymbol{y} \left( \sigma_i -\dfrac{1}{2c^2} \sigma \partial_i \partial_t X \right) + {\rm O}(c^{-4}),
\end{equation}
where  the super-potential $X\left(t,\boldsymbol{y}\right)$ has   been introduced in Eq. (\ref{eq:X-field-torsion}). By exploiting  Eq. (\ref{eq:nabla_tilde_T}), we can demonstrate that 
\begin{eqnarray}
\dfrac{{\rm d}}{{\rm d}t} I\left(t\right) &=&0, \\
\dfrac{{\rm d}}{{\rm d}t} \mathscr{P}_i\left(t\right) =\dfrac{{\rm d}^2}{{\rm d}t^2} I_i\left(t\right) &=&0, \\
\dfrac{{\rm d}}{{\rm d}t} J_i\left(t\right) &=&0.
\end{eqnarray}

The PN expansion, up to ${\rm O}(c^{-2},c^{-3},c^{-2})$ terms, of the  total stress-energy pseudo-tensor (\ref{eq:tilde-tau-alpha-beta-def-1}) yields 
\begin{eqnarray}
\tilde{\mathfrak{T}}^{00} &=& \left(-g^{\rm in}\right) \Theta^{00} -\dfrac{7}{8 \pi G} \partial_i U^{\rm in} \partial_i U^{\rm in} +{\rm O}(c^{-2}), \label{eq:PN-expansion-of-tilde-Tau-00}
\\
\tilde{\mathfrak{T}}^{0i} &=& \left(-g^{\rm in}\right) \Theta^{0i} +\dfrac{1}{4 \pi G c} \left( 3\partial_t U^{\rm in} \partial_i U^{\rm in} + 8 \partial_k U^{\rm in}  \partial_{[i} U^{\rm in}_{k]}  \right) 
\nonumber \\
&+&{\rm O}(c^{-3}),\label{eq:PN-expansion-of-tilde-Tau-0i}
\\
\tilde{\mathfrak{T}}^{ij} &=& \left(-g^{\rm in}\right) \Theta^{ij} +\dfrac{1}{4 \pi G}\left( \partial_i U^{\rm in} \partial_j U^{\rm in}-\dfrac{1}{2}\delta_{ij} \partial_k U^{\rm in} \partial_k U^{\rm in}\right)
\nonumber \\
&+& {\rm O}(c^{-2}),\label{eq:PN-expansion-of-tilde-Tau-ij}
\end{eqnarray}
where the determinant of the inner metric field reads as
\begin{equation}\label{eq:PN-expansion-determinant}
    g^{\rm in} = -\left(1 + \dfrac{4}{c^2} U^{\rm in} \right)+ {\rm O} \left(c^{-4}\right).
\end{equation}
Bearing in mind Eqs. (\ref{eq:PN-expansion-of-tilde-Tau-00})--(\ref{eq:PN-expansion-determinant}), it is simple to show that  Eqs. (\ref{eq:I-expression-torsion})--(\ref{eq:P-i-expression-torsion}) can be equivalently written  as
\begin{eqnarray}
I\left(t\right) &=& \dfrac{1}{c^2} \int {\rm d}^3 \boldsymbol{y} \, \tilde{\mathfrak{T}}^{00}, \label{eq:I-torsion-tilde-tau}
\\
I_i\left(t\right) &=& \dfrac{1}{c^2} \int {\rm d}^3 \boldsymbol{y} \,y_i \tilde{\mathfrak{T}}^{00}, 
\\
\mathscr{P}_i\left(t\right) &=& \dfrac{1}{c} \int {\rm d}^3 \boldsymbol{y} \, \tilde{\mathfrak{T}}^{0i},
\label{eq:P-i-torsion-tilde-tau}
\\
J_i\left(t\right) &=& \dfrac{1}{c} \int {\rm d}^3 \boldsymbol{y} \, \epsilon_{ijk} y_j \tilde{\mathfrak{T}}^{0k}, \label{eq:J-i-torsion-tilde-tau}
\end{eqnarray}
where the evaluation of $J_i\left(t\right)$ necessitates only ${\rm O}(c^{-1})$ accuracy in Eq. (\ref{eq:PN-expansion-of-tilde-Tau-0i}).

In the following, we will analyse $I$, $\mathscr{P}_i$ and $J_i$ in two ways: in Sec. \ref{Sec:Flat-space-limit} we will evaluate them in flat spacetime; in Sec. \ref{Sec:surface-integrals} we will consider surface integrals. Both arguments permit to clearly disclose their physical meaning.

\subsubsection{Flat-space limit} \label{Sec:Flat-space-limit}
We can  evaluate Eqs. (\ref{eq:I-torsion-tilde-tau}), (\ref{eq:P-i-torsion-tilde-tau}) and (\ref{eq:J-i-torsion-tilde-tau}) in the absence of the torsion and the gravitational field in two steps: first, we calculate their expressions in GR (i.e., vanishing torsion) and after that in  Minkowski spacetime (i.e., no gravity). This means that we need a \emph{modus operandi} to \qm{detach} the spin of the matter field from the torsion (i.e., the geometry) in such a way that we can  \qm{switch off} the latter while maintaining the former. 

The GR limit of  Eqs. (\ref{eq:I-torsion-tilde-tau}), (\ref{eq:P-i-torsion-tilde-tau}) and (\ref{eq:J-i-torsion-tilde-tau}) can be taken by replacing, in the formula of $\tilde{\mathfrak{T}}^{\alpha \beta}$ (cf. Eq. (\ref{eq:tilde-tau-alpha-beta-def-1})), $\Theta^{\mu \nu}$ with the GR metric stress-energy tensor, which we hereafter denote with $\hat{T}^{\mu \nu}$ to distinguish it from the one defined in EC theory (see  Eq. (\ref{eq:metric_stress-energy_tensor})). This procedure turns out  to be equivalent to performing in Eqs. (\ref{eq:I-torsion-tilde-tau}), (\ref{eq:P-i-torsion-tilde-tau}), and (\ref{eq:J-i-torsion-tilde-tau}) the following substitution (cf. Eqs. (\ref{eq:mu_tensor}), (\ref{eq:canonical_stress-energy_tensor}), and (\ref{eq:tilde-T-alpha-beta}))
\begin{align}\label{eq:substitution-torsion-1}
   & \mathbb{T}^{\mu \nu} -\overstarbf{\nabla}_{\lambda}\left(\tau^{\mu \nu \lambda} + \tau^{\lambda \mu \nu} + \tau^{\lambda \nu \mu}\right) + \dfrac{\chi}{2} \mathcal{S}^{\mu \nu} 
   \nonumber \\
  & \longrightarrow \; \hat{\mathbb{T}}^{\mu \nu} -\hat{\nabla}_{\lambda}\left(\hat{\tau}^{\mu \nu \lambda} + \hat{\tau}^{\lambda \mu \nu} + \hat{\tau}^{\lambda \nu \mu}\right),
\end{align}
where $\hat{\mathbb{T}}^{\mu \nu}$ and $\hat{\tau}^{\mu \nu \lambda}$ are the GR canonical stress-energy tensor and the spin density tensor, respectively, which  differ, in general, from the corresponding EC definitions  (\ref{eq:canonical_spin_tensor}) and (\ref{eq:definition-of-canonical-stress-energy-tensor}).  In GR we can  adopt the canonical definition of spin density tensor \cite{Hehl1976_tensor}
\begin{equation}
    \sqrt{-g} \, \hat{\tau}^{ \mu \nu \lambda} = \dfrac{\partial \hat{\mathcal{L}}_{\rm m}}{\partial \left( \partial_\lambda \Psi \right)} f^{[\nu \mu]} \Psi,
\end{equation}
$\Psi$ being the matter field, $\hat{\mathcal{L}}_{\rm m}$  the matter Lagrangian density, minimally coupled to the gravitational field through the  replacements $\partial_{\mu} \Psi \rightarrow \hat{\nabla}_{\mu} \Psi$ and $\eta_{\mu \nu} \rightarrow g_{\mu \nu}$,  and $f^{\nu \mu}$  the matrices representing the generators of an infinitesimal coordinate transformation appropriate to the representation of $\Psi$.  

By means of the prescription (\ref{eq:substitution-torsion-1}), Eqs. (\ref{eq:I-torsion-tilde-tau}), (\ref{eq:P-i-torsion-tilde-tau}), and (\ref{eq:J-i-torsion-tilde-tau}) go over into their GR counterparts $\hat{I}$, $\hat{\mathscr{P}}_i$, and $\hat{J}_i$, i.e.,
\begin{align} \label{eq:hat-I-hat-I-i-hat-J-i} 
   &I\qquad \longrightarrow\quad \hat{I} =\dfrac{1}{c^2} \int {\rm d}^3 \boldsymbol{y} \, \mathfrak{T}^{00},
   \nonumber \\
   &\mathscr{P}_i\ \ \  \longrightarrow\ \hat{\mathscr{P}}_i = \dfrac{1}{c} \int {\rm d}^3 \boldsymbol{y} \, \mathfrak{T}^{0i}, 
   \nonumber \\
   &J_i\ \ \ \ \longrightarrow\ \ \ \hat{J}_i = \dfrac{1}{c} \int {\rm d}^3 \boldsymbol{y} \, \epsilon_{ijk} y_j \mathfrak{T}^{0k},
\end{align}
where $\mathfrak{T}^{\alpha \beta}$ can be read from Eq. (\ref{eq:tau-alpha-beta}) (recall that, in the notations adopted in this section, the metric stress-energy tensor is indicated with $\hat{T}^{\alpha \beta}$).  Therefore, we can conclude that Eqs. (\ref{eq:I-torsion-tilde-tau}), (\ref{eq:P-i-torsion-tilde-tau}) and (\ref{eq:J-i-torsion-tilde-tau}) reduce, in flat spacetime, to 
\begin{eqnarray}
I^{{\rm flat}} &=& \dfrac{1}{c^2} \int {\rm d}^3 \boldsymbol{y} \, \mathbb{T}_{{\rm flat}}^{00} =\dfrac{1}{c^2} \int {\rm d}^3 \boldsymbol{y} \, T_{{\rm flat}}^{00}, \label{eq:I-flat}
\\
\mathscr{P}^{{\rm flat}}_i &=& \dfrac{1}{c} \int {\rm d}^3 \boldsymbol{y} \, \mathbb{T}_{{\rm flat}}^{i0}=\dfrac{1}{c} \int {\rm d}^3 \boldsymbol{y} \, T_{{\rm flat}}^{i0}, \label{eq:P-i-flat}
\\
J^{{\rm flat}}_i &=& \dfrac{1}{c} \int {\rm d}^3 \boldsymbol{y} \, \epsilon_{ijk} \left(y_j \mathbb{T}_{{\rm flat}}^{k0} + \tau_{{\rm flat}}^{jk0}\right)
\nonumber \\
&=&\dfrac{1}{c} \int {\rm d}^3 \boldsymbol{y} \, \epsilon_{ijk} y_j T^{k0}_{{\rm flat}}, 
\label{eq:J-i-flat}
\end{eqnarray}
where
\begin{equation}
    T^{\mu \nu}_{{\rm flat}} = \mathbb{T}^{\mu \nu}_{{\rm flat}} -\partial_\lambda\left(\tau_{{\rm flat}}^{\mu \nu \lambda} + \tau_{{\rm flat}}^{\lambda \mu \nu} + \tau_{{\rm flat}}^{\lambda \nu \mu}\right),
\end{equation}
is the symmetric stress-energy tensor associated to the canonical stress-energy tensor $\mathbb{T}_{{\rm flat}}^{\mu \nu}$ according to the \emph{Belinfante-Rosenfeld symmetrization procedure} \cite{Davis1970}. 

Equations (\ref{eq:I-flat})--(\ref{eq:J-i-flat}) agree with the special-relativity limit of Eqs. (\ref{eq:Mathisson-Papapertrou_parameters})  and (\ref{eq:Mathisson-Papapertrou_parameters-spin}). In particular, $I^{{\rm flat}}$ and $\mathscr{P}^{{\rm flat}}_i$ give the mass and the three-momentum of the source, respectively, while  $J^{{\rm flat}}_i$  the total angular momentum of the source \cite{Maiani2016,Gasperini-DeSabbata,Hehl1976_tensor}. Therefore, our PN-expanded results (\ref{eq:I-torsion-tilde-tau}), (\ref{eq:P-i-torsion-tilde-tau}) and (\ref{eq:J-i-torsion-tilde-tau}) lead to well-defined  objects in flat spacetime.  For this reason, we can conclude that also in EC model, like in Einstein theory, the 1PN lowest-order source multipole moments have a precise physical content and Eqs. (\ref{eq:I-expression-torsion}), (\ref{eq:P-i-expression-torsion}), and (\ref{eq:J-i-expression-torsion}) give rise to a generalized notion of total Arnowitt-Deser-Misner (ADM) \emph{mass} and  \emph{three-momentum}, and  total \emph{angular momentum} of the system, respectively.  

\subsubsection{Surface integrals} \label{Sec:surface-integrals}
At 1PN level, the contributions of $t^{\alpha \beta}_{\rm H}$ to Eqs. (\ref{eq:I-torsion-tilde-tau})--(\ref{eq:J-i-torsion-tilde-tau})  can be ignored since
\begin{equation}
\left(-g^{\rm in}\right)t^{\alpha \beta}_{\rm H} = {\rm O}\left(c^{-2},c^{-3},c^{-4}\right), 
\end{equation}
(the form of $t^{\alpha \beta}_{\rm H}$ is reported in Eq. (\ref{t-alpha-beta-H-1-GR}), see also the discussion before Eq. (\ref{eq:math-problem-EC2})). This allows  to evaluate Eqs. (\ref{eq:I-torsion-tilde-tau}) and (\ref{eq:P-i-torsion-tilde-tau}) in terms of Gaussian flux integrals over a two-dimensional closed surface $\mathscr{S}$, which completely surrounds the source and lies in a  three-surface of constant time $x^0$. From Eq. (\ref{eq:I-torsion-tilde-tau}), we have
\begin{equation} \label{eq:surface-I}
     I= \dfrac{1}{c^2} \int {\rm d}^3 \boldsymbol{y} \, \tilde{\mathfrak{T}}^{00} = \dfrac{1}{\chi c^2  } \oint_{\mathscr{S}} \partial_j H^{0j0k} {\rm d}S_k + {\rm O}\left(c^{-4}\right),
 \end{equation}
where 
 \begin{equation}
     H^{\mu \alpha \nu \rho} = \mathfrak{g}^{\mu \nu}\mathfrak{g}^{\alpha \rho} - \mathfrak{g}^{\mu \rho} \mathfrak{g}^{\alpha \nu}.
\end{equation}

Along the same lines, we can write for Eq. (\ref{eq:P-i-torsion-tilde-tau})
\begin{align} \label{eq:surface-P-i}
     \mathscr{P}_i &= \dfrac{1}{c} \int {\rm d}^3 \boldsymbol{y} \, \tilde{\mathfrak{T}}^{0i} =     \dfrac{1}{\chi c } \oint _{\mathscr{S}} \partial_\mu H^{i \mu 0 k} {\rm d}S_k + {\rm O}\left(c^{-4}\right).
 \end{align}
Equations (\ref{eq:surface-I}) and (\ref{eq:surface-P-i}) display  an expression resembling formally the ADM mass and three-momentum, respectively, of an asymptotically flat spacetime in GR \cite{MTW,Poisson-Will2014}. Since the metric tensor components occurring in Eqs. (\ref{eq:surface-I}) and (\ref{eq:surface-P-i}) contain also torsion contributions, we  can conclude that the four-vector $\mathscr{P}^{\mu} = (c I, \mathscr{P}^i)$, having components given by Eqs. (\ref{eq:I-expression-torsion}) and (\ref{eq:P-i-expression-torsion}),  represents a generalized ADM four-momentum in EC theory. In particular,   the torsion tensor introduces in Eq. (\ref{eq:I-expression-torsion}) corrections going like the square of the energy-dipole-moment density (cf. Eq. (\ref{eq: tau_physical})). We also note that, in our approach, the volume integrals, and hence also the surface integrals, defining $I$ and  $\mathscr{P}_i$ can be evaluated with the near-zone information available to us. 

It is known that  in GR the ADM four-momentum has a well-defined meaning in the asymptotically flat region outside the source, where linearized GR  guarantees that it behaves as a special relativistic four-vector under Lorentz transformations, and  that it is invariant under infinitesimal coordinate transformations \cite{MTW}. This is  true also in  EC model and  we  can calculate, in particular, the integral $\oint_{\mathscr{S}} \partial_j H^{0j0k} {\rm d}S_k $ within the linearized EC theory (see Sec. \ref{sec:Linearized-EC}). In fact,  since the two-surface $\mathscr{S}$ resides in the asymptotically flat region of the spacetime $\mathscr{M}$, it can be evaluated by means of Eq. (\ref{eq:linearizedEC-metric}). In this way, we obtain, in the weak-field region far from the source,
 \begin{align} \label{eq:surface-I-2}
  \dfrac{1}{\chi c^2  } \oint_{\mathscr{S}} \partial_j H^{0j0k} {\rm d}S_k  &= \dfrac{1}{\chi c^2  } \oint_{\mathscr{S}} \left(\partial_j g_{jk} - \partial_k g_{jj}\right){\rm d}S_k.
\end{align}
Having recovered a quantity having the same functional form of the ADM mass in GR \cite{MTW,Poisson-Will2014}, we can interpret it  as generalized ADM mass of EC theory,  confirming thus what we have obtained before. 

Finally, $J_i$, given by Eq. (\ref{eq:J-i-torsion-tilde-tau}), cannot be written via a surface integral, because it is known modulo ${\rm O}(c^{-2})$ terms (see Eq. (\ref{eq:J-i-expression-torsion})). Explicitly, it can be written as (cf. Eqs. (\ref{eq:S_expansion_2}) and (\ref{eq:sigma_i}))
 \begin{equation}
     J_i = \dfrac{1}{c} \int {\rm d}^3 \boldsymbol{y} \, \epsilon_{ijk} y_j T^{0i} + {\rm O}(c^{-2}),
 \end{equation}
and we see that it is  the EC generalization of Eq. (\ref{eq:J-i-flat}). 

\subsection{The asymptotic gravitational  waveform}
\label{Sec:asymptotic-waveform}
Having obtained the sought-after relation linking  the source multipole moments and the radiative moments (cf. Eq. (\ref{eq:radiative-source_1PN})), we can evaluate the expression of the asymptotic waveform  at 1PN level. 
Let 
\begin{equation}\label{eq:metric-radiative-coordinates}
    \mathscr{G}^{\rm ext}_{\mu \nu}(X) = \eta_{\mu \nu} + \mathscr{H}_{\mu \nu}(X),
\end{equation}
denote the external metric written in radiative coordinates $X^\mu=(cT,\boldsymbol{X})$.  The asymptotic gravitational waveform of the PN source is defined starting from the transverse-traceless (TT) projection of the leading $R^{-1}$ term of the far-zone expansion at future null infinity (i.e., $R \to \infty$ with $\mathcal{U} \equiv T-R/c$ and $\boldsymbol{N} \equiv \boldsymbol{X}/R$ fixed) of the  metric coefficients (\ref{eq:metric-radiative-coordinates}) (for details, we refer the reader to \cite{Blanchet2014,Maggiore:GWs_Vol1}).  At 1PN order, one finds that this quantity  reads as  \cite{Blanchet-Damour1989,Blanchet1995}
\begin{align} \label{eq:gravitational_wave_amplitude}
    \mathscr{H}_{ij}^{\rm TT}(X^\mu) & = \dfrac{2G}{c^4 R} \mathcal{P}_{ijkl}(\boldsymbol{N})  \Biggr\{ U_{kl}(\mathcal{U}) 
    \nonumber \\
    & + \dfrac{1}{c} \left[  \dfrac{1}{3} N_a U_{kla}(\mathcal{U}) +\dfrac{4}{3} \epsilon_{ab(k} V_{l)a}(\mathcal{U}) N_b \right]
    \nonumber \\
    & +\dfrac{1}{c^2} \left[\dfrac{1}{12}N_{ab} U_{klab} (\mathcal{U})
     + \dfrac{1}{2} \epsilon_{ab(k} V_{l)ac}(\mathcal{U}) N_{bc} \right]
     \nonumber \\
     & + {\rm O}(c^{-3}) \Biggr\},
\end{align}
where $\mathcal{P}_{ijkl}(\boldsymbol{N})$ is the TT projection operator onto the plane orthogonal to $\boldsymbol{N}$, i.e., 
\begin{align}
    \mathcal{P}_{ijkl} (\boldsymbol{N}) & \equiv \mathcal{P}_{ik}\mathcal{P}_{jl}-\dfrac{1}{2}\mathcal{P}_{ij}\mathcal{P}_{kl},
\end{align}
where $\mathcal{P}_{ij} (\boldsymbol{N})  \equiv \delta_{ij}-N_i N_j$.  At this level, in order to compute (\ref{eq:gravitational_wave_amplitude}), we only need 1PN accuracy for the  mass-type radiative quadrupole moment $U_{kl}$, whereas   for  the remaining radiative moments their 0PN expression    suffices. From   Eqs. (\ref{eq:I_L_torsion}) and (\ref{eq:radiative-source_1PN}), we find that
\begin{align}  \label{eq:radiative-U-ij}
U_{kl}(u) & = \dfrac{{\rm d}^2}{{\rm d}u^2} \Biggl\{ \int {\rm d}^3 \boldsymbol{y} \, y_{<kl>} \sigma(\boldsymbol{y},u) 
\nonumber \\
&+ \dfrac{1}{14}\dfrac{1}{c^2} \dfrac{{\rm d}^2}{{\rm d}u^2} \int {\rm d}^3 \boldsymbol{y} \, y_{<kl>} \,  \boldsymbol{y}^2 \sigma(\boldsymbol{y},u) 
\nonumber \\
& - \dfrac{20}{21}\dfrac{1}{c^2} \dfrac{{\rm d}}{{\rm d}u} \int {\rm d}^3 \boldsymbol{y} \, y_{<kli>} \sigma_i (\boldsymbol{y},u)\Biggr\}
\nonumber \\
& + {\rm O}(c^{-3}),
\end{align}
where we recall that $\sigma$ and $\sigma_i$ are given by Eqs. (\ref{eq:sigma_tilde}) and (\ref{eq:sigma_i}), respectively. Then, the spin contributions to  Eq. (\ref{eq:radiative-U-ij}) appear implicitly in $T^{\mu \nu}$, and explicitly, via the function $\sigma(\boldsymbol{y},u) $,  in the terms ${}^{(0)}\mathcal{S}^{00}$ and ${}^{(0)}\mathcal{S}^{jj}$ (cf. Eqs. (\ref{eq:S_expansion}) and (\ref{eq:S_expansion_2})). The latter factors, bearing in mind Eqs. (\ref{Eq:S_alphabeta}) and (\ref{eq: tau_physical})--(\ref{eq:tau_PN_expansion}), introduce in the waveform (\ref{eq:gravitational_wave_amplitude}) corrections proportional  to the square of the energy-dipole-moment density of the system. 

In GR, the lowest-order SO effects introduced by the classic angular momentum in the source multipole moments emerge at  1.5PN order in the case of the mass-type multipole moments $I_L$ and at 0.5PN level for the current-type multipole moments $J_L$ \cite{Kidde1993,Blanchet2014}. On the other hand, in EC theory, the above analysis shows that  the \emph{explicit} contributions of the spin to the source multipole moments, and hence also to the radiative multipole moments, occur  only for the 1PN mass-type  multipole moments, whereas in the current-type  ones, for which we have only Newtonian accuracy, spin modifications can  \emph{implicitly} be contained in $T^{0i}$ (due to the presence of  $\sigma_i$, see Eq. (\ref{eq:J_L_torsion})).  As pointed out before, this   means that in the 1PN waveform (\ref{eq:gravitational_wave_amplitude}) the tensor $\mathcal{S}^{\mu \nu}$ yields
 explicit  spin corrections  to the leading  $U_{kl}(\mathcal{U})$ term. 

The evaluation of the waveform (\ref{eq:gravitational_wave_amplitude}) necessitates the knowledge of the source dynamics at 1PN. The explicit form of the dynamical equations can be obtained from the conservation laws (\ref{eq:conservation-law-energy-momentum}) and (\ref{eq:conservation-law-angular-momentum}) once the specific expressions for the canonical energy-momentum and spin tensors (Eqs. (\ref{eq:definition-of-canonical-stress-energy-tensor}) and (\ref{eq:canonical_spin_tensor}), respectively) are provided. In other words,   a particular model characterizing
 the matter source should be assigned in order to investigate its dynamics.  However, the analysis of Eq. (\ref{eq:Mathisson-Papapertrou_eq}), which describes  the translational motion in the case of  a spinning test body, reveals precious information regarding the source dynamics in EC theory. First of all, the interaction between spin and curvature, represented by the combination $R^{\mu}_{\phantom{\mu}\nu \rho \sigma} S^{\rho \sigma} u^{\nu}$ on the right-hand side of Eq. (\ref{eq:Mathisson-Papapertrou_eq}), has a completely  different nature from the corresponding GR expression occurring in the Mathisson-Papapetrou equation \cite{Papapetrou1951}. Indeed, in GR this term is obtained in  second or dipole-particle approximation, whereas in EC model it arises already  in first approximation (i.e., the pole-particle approximation). Unlike the GR case, where it involves the macroscopic angular momentum of the body,  in EC theory this factor  depends on the 
 intrinsic spin of the test particle and  hence it vanishes  by taking the formal limit $\hbar \to 0$ \cite{Hehl1971}. Furthermore, due to the presence of the Riemann tensor,  it  contains  contributions proportional to  the torsion of the background geometry (cf. Eqs. (\ref{eq:torsion_tensor}),  (\ref{eq:symmetric_part}), and (\ref{eq:EC-Riemann-tensor})). Therefore, the spin of the test particle will interact with the spin of the matter source generating the background gravitational field. The torsion of the background geometry also contributes to  the quantity  $K^{\mu}_{\phantom{\mu} \nu \lambda} P^{[\lambda} u^{\nu]}$ appearing in the left-hand side of Eq. (\ref{eq:Mathisson-Papapertrou_eq}), which therefore yields interaction terms similar to those described before if we recall that   four-momentum $P^\mu$ and the test particle four-velocity $u^\mu$ are no longer related through the usual proportionality relation involving the test particle rest mass. 
 
Finally, from the  conservation law (\ref{eq:conservation-law-angular-momentum}), we obtain the equations controlling the rotational degrees of freedom of the source. Also in this case we expect 
interaction terms similar to those underlying  the translational dynamics. 

A practical application of the above results deserves consideration in a separate paper.

\section{Conclusions} 
\label{Sec:conclusion}
The GWs generation theory accomplishes several goals, like: acquiring further (and sometimes peculiar) information about existing and new gravitational systems, revealing the fascinating nature and proprieties of BHs, inquiring the physics of dense matter in neutron stars, shading light about the early stages of the Universe formation. Encouraged by the recent discoveries and the great amount of actual and near-future very sensitive data, which will offer viable opportunities to scrape the quantum world, it becomes fundamental to update our theoretical assessments. In this perspective, we considered interesting to explore the GWs generation problem in the context of EC theory at 1PN order by employing the Blanchet-Damour formalism, briefly recalled, in the GR case, in Sec.  \ref{sec:problem}.

In EC theory (see Sec. \ref{sec:GW_EC_theory}) the initial well-posed mathematical problem (\ref{eq:proc-EC}) is in general very complex. We simplify the mathematical scheme by assuming that $S^{\alpha\mu}{}_\mu=0$,  an assumption widely employed in different physical contexts which permits to obtain the following advantages: (1) the EC field equations become more treatable; (2) the gauge conditions in the interior and exterior zones are the same, avoiding thus junction conditions of two different (i.e., internal and external) coordinate systems. In addition, the compactness of the spin density tensor $\tau^{\alpha\beta\lambda}$  guarantees that the EC field equations reduce to those of GR theory in the exterior zone and thus the assumption of weak torsion field is unnecessary. For negligibly self-gravitating sources, the resolution method can be tackled by considering the linearized EC theory. Instead, for weakly self-gravitating sources, the Blanchet-Damour formalism is applied. Equations (\ref{eq:radiative-canonical_1PN}) and (\ref{eq:radiative-source_1PN}) represent the concluding result of the matching procedure and the solution of the GWs generation problem in EC model at 1PN level. This section concludes with   the analysis about the physical meaning of the lowest-order source multipole moments and the 1PN asymptotic gravitational waveform in EC theory. In particular, the critical argument behind the former  investigation relies on the philosophy that through the disentanglement of the spin from the torsion tensor (i.e., its geometrical counterpart), it is possible to descend to the GR theory keeping the spin contributions, which are not anymore coupled to the geometrical background. 

As concluding discussion, we estimate the order of magnitude of the spin contributions to the GW signal in EC theory, in order to make a comparison with respect to GR. To this end, we consider the following linearized metric stress-energy tensor, defined in the compact region $\Omega$, for a spinning and pressureless fluid:
\begin{equation} \label{eq:SETGEN}
T^{\alpha \beta} =\rho \dfrac{u^\alpha u^\beta}{c^2} 
           -\Phi^{\alpha\beta}(s^{\mu\nu},u^\lambda),
\end{equation}
where $\Phi^{\alpha\beta}$ has the dimension of the time-variation of an angular momentum over a volume, and $s^{\mu\nu}$ is the spin density tensor. We underline that $T^{\alpha\beta}$ is a generic tensor, which does not refer to any particular physical model, because we aim at providing a \emph{model-independent estimate of the EC effects}. We consider the following assumptions: the energy density is $\rho=\mathcal{M} c^2/\mathcal{V}$, where $\mathcal{M}$ and $\mathcal{V}={\rm Vol}(\Omega)$ are the mass and the volume of the whole gravitational source, respectively; $\Phi^{\alpha\beta}=2u^{(\beta}\partial_\mu s^{\alpha)\mu}$; the velocity $u^\alpha$ is constant in time and space.

\begin{figure*}[t!]
    \centering
    \includegraphics[scale=0.62]{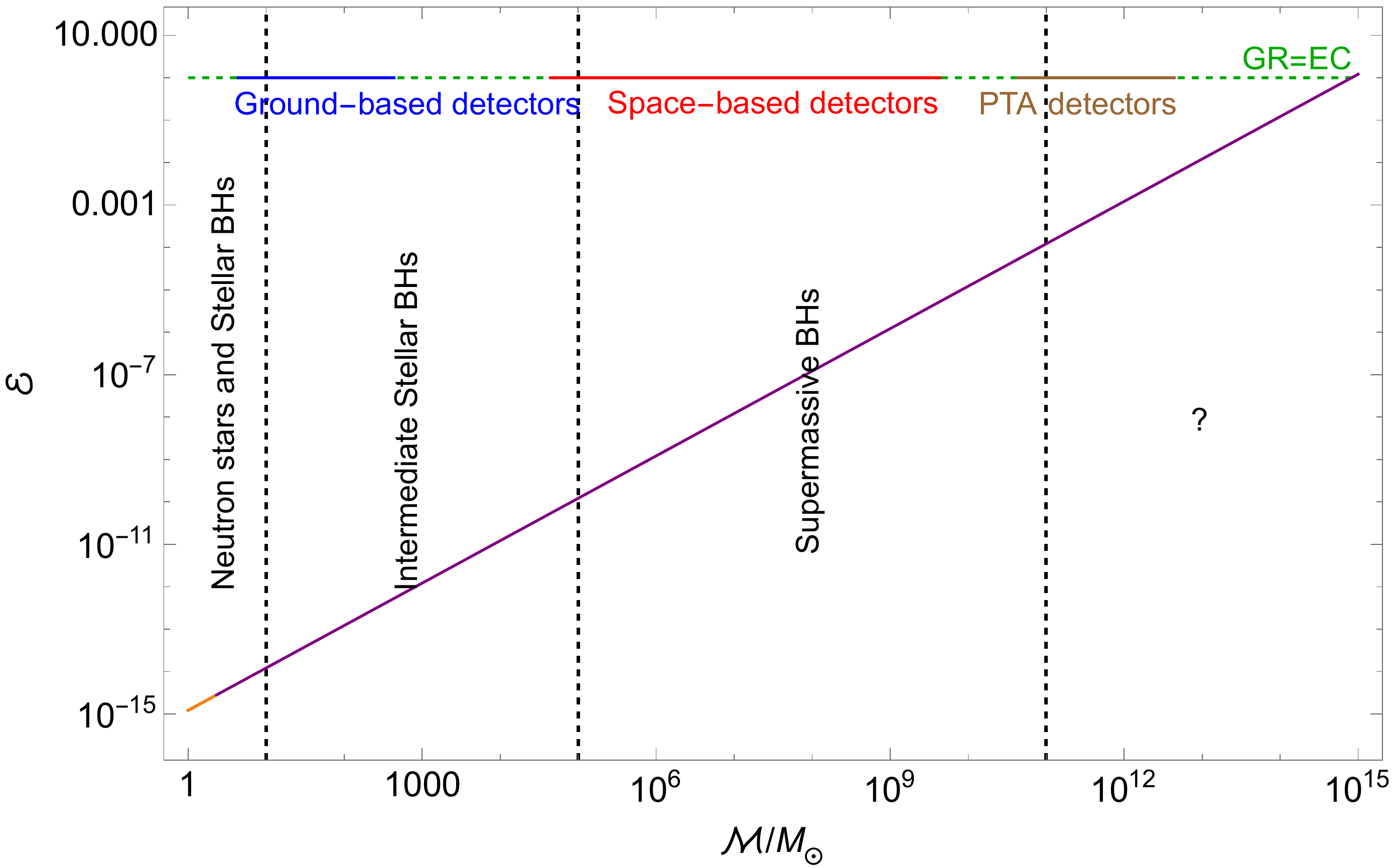}
    \caption{Relative EC spin contribution over the GW signal in GR, $\mathcal{E}$, in terms of the number of solar masses of the whole gravitational system, $\mathcal{M}/M_\odot$. The purple line describes the BH case, whereas the orange one characterises the neutron stars. The dashed vertical lines delimit the neutron stars ($1-2.14\ M_\odot$) and stellar BHs ($1-10\ M_\odot$), intermediate stellar BHs ($10-10^5\ M_\odot$), and supermassive BHs ($10^5-10^{11}\ M_\odot$) ranges. {\bf No astrophysical objects with masses greater than $10^{11} \ M_\odot$ have been detected so far.} The dashed green line is at $\mathcal{E}=1$, where EC and GR effects are comparable. The blue, red, and brown lines are the observational mass ranges of ground-based, space-based, and PTA  detectors, respectively. In our estimate, the purple line will reach the green dashed line at $\mathcal{M}=8.23\times10^{14}\ M_\odot$, which corresponds to a frequency of the order of $ 5\times10^{-12}$ Hz (cf. Eq. (\ref{eq:fGW-ISCO})).}
    \label{fig:Fig2}
\end{figure*}

The linearized problem $\Box \theta^{\alpha\beta}=-\chi T^{\alpha\beta}$ (cf. Eq. (\ref{eq:linearized_Einstein-Cartan_compact})), can be further approximated to $\Delta \theta^{\alpha\beta}=-\chi T^{\alpha\beta}$, whose solution is expressed by the following function
\begin{align} \label{eq:solsim}
\theta^{\alpha\beta}&=\frac{4G}{c^4 \mathcal{R}}\int_{\Omega} T^{\alpha\beta} {\rm d}^3\boldsymbol{x}\notag\\
&=\frac{4G}{c^4 \mathcal{R}}\left[\mathcal{M} u^\alpha u^\beta-2u^{(\beta}\int_{\Omega}\partial_i s^{\alpha) i}\ {\rm d}^3\boldsymbol{x}\right],
\end{align}
where we have assumed that $s^{\alpha \mu}$ is independent of time, and $\mathcal{R}$ is the distance between the observer and the source, supposed to be located very far, which permits to move $\mathcal{R}$ out of the integral. Regarding the integral in Eq. (\ref{eq:solsim}), we can now apply the Gauss theorem and do the following approximation 
\begin{equation}
\int_{\Omega}(\partial_i s^{\alpha i}) {\rm d}^3\boldsymbol{x}=\int_{\partial\Omega} (s^{\alpha i} \hat{\nu}_i) {\rm d}\Sigma\approx s^{\alpha} \Sigma,  
\end{equation}
where $\Sigma$ is the measure of the boundary of $\Omega$ calculated as $\Sigma=8\pi (2G\mathcal{M}/c^2)^2$ (namely the sum of the surfaces of the two compact objects in the binary system modelled as spheres), and $s^{\alpha}$, which has the dimension of a spin density, is obtained by  projecting $s^{\alpha \beta}$ along the direction where the spins in the matter are aligned. We further assume: $|\boldsymbol{u}|\equiv\sqrt{\delta_{ij}u^iu^j}=10^4\ \mbox{m/s}$ is the spatial velocity of a compact binary system; $s^\alpha=s \hat{s}^\alpha$ with $s\equiv \sqrt{s^\alpha s_\alpha}=n\hbar$, where $n=10^{44}$ m${}^{-3}$ is estimated as the inverse of the nucleon volume (supposing that all nucleons in a given volume have aligned spins), and $\hat{s}^\alpha$ stands for the versor of $s^\alpha$.  Finally, we obtain
\begin{equation}
\theta^{\alpha\beta}=\frac{4G}{c^4 \mathcal{R}}\left[\mathcal{M}u^\alpha u^\beta-2n\hbar\hat{s}^{(\alpha}u^{\beta)}\Sigma\right].
\end{equation}
Assuming that both $u^\alpha$ and $s^\alpha$ have  non-vanishing components only along the $z-$direction, we can write $u^i=u^z=|\boldsymbol{u}|$ and $s^i=s^z=s\hat{s}^z$. We distinguish the two gravitational amplitudes:
\begin{eqnarray}
\theta_{zz}^{{\rm GR}}&=&\frac{4G}{c^4 \mathcal{R}}\left(\mathcal{M}|\boldsymbol{u}|^2\right),\\
\theta_{zz}^{{\rm EC}}&=&\frac{4G}{c^4 \mathcal{R}}\left(\mathcal{M}|\boldsymbol{u}|^2-n\hbar |\boldsymbol{u}|\Sigma\right),
\end{eqnarray}
where the former is the solution within GR, whereas the latter is framed in EC theory. This allows to finally compute the spin contributions over the GW signal in GR theory through the following formula:
\begin{equation} \label{eq:ERRREL}
\mathcal{E}\equiv\frac{|\theta_{zz}^{{\rm EC}}-\theta_{zz}^{{\rm GR}}|}{|\theta_{zz}^{{\rm GR}}|}=\frac{n\hbar\Sigma}{\mathcal{M}|\boldsymbol{u}|}.  
\end{equation}
In Fig. \ref{fig:Fig2}, we plot $\mathcal{E}$ in terms of numbers of solar masses $\mathcal{M}/M_\odot$ (namely the total mass of the system). The corresponding GW frequency $f_{\rm GW}$ can be roughly  estimated via the formula \cite{Maggiore:GWs_Vol1}
\begin{equation} \label{eq:fGW}
 f_{\rm GW}=   \dfrac{1}{\pi} \dfrac{\sqrt{G \mathcal{M}}}{\mathscr{R}\sqrt{\mathscr{R}}},
\end{equation}
where $\mathscr{R}$ is a generic orbital radius. To fix the ideas, in Fig. \ref{fig:Fig2} we have considered  the Schwarzschild innermost stable circular orbit (ISCO) radius
\begin{equation}
    \mathscr{R}_{\rm ISCO} = \dfrac{6 G \mathcal{M}}{c^2},
\end{equation}
for which Eq. (\ref{eq:fGW}) becomes 
\begin{equation} \label{eq:fGW-ISCO}
 f_{\rm GW}^{\rm ISCO} \simeq 4.4 \,{\rm kHz}\; (M_\odot/\mathcal{M}). 
\end{equation}
In addition, we have  shown in correspondence with the horizontal line $\mathcal{E}=1$ (where GR and EC effects are of the same order) the frequency ranges  for ground-based ($10-10^3$ Hz), space-based ($10^{-6}-10^{-1}$ Hz), and PTA ($10^{-9}-10^{-7}$ Hz) detectors in terms of $\mathcal{M}/M_\odot$. In principle, the horizontal line of the frequency ranges can be  extended  for  values  $\mathcal{E} \lesssim 1$. However,   this depends  both on the specific model exploited to describe the EC effects and on the GR template used for the  detection of the GW frequency. Another important information on EC model is encoded in the values attained by $\mathcal{E}$ for the different classes of plotted astrophysical objects, which read as follows: $\mathcal{E}\in[10^{-15},3\times10^{-15}]$ for neutron stars, $\mathcal{E}\in[10^{-15},10^{-14}]$ for stellar BHs, $\mathcal{E}\in[10^{-15},10^{-10}]$ for intermediate stellar BHs, and $\mathcal{E}\in[10^{-10},10^{-4}]$ for supermassive BHs.   

From our speculative graph, we find that for masses  $\mathcal{M} \ge8.23\times10^{14}\ M_\odot$ the EC effects are of the same order of magnitude as in GR. This corresponds to GW frequencies of the order of $5\times 10^{-12}$ Hz, which are three orders of magnitude lower than  the minimum frequency range of PTA \emph{apparati}. Since this value corresponds to an orbital period of the order of $6 \times 10^3$ years, our speculative analysis can explain why quantum effects have not been detected to date with the current technology. We also note that  BHs with masses grater than $10^{11}\ M_\odot$ have never been observed. In addition, the detection of a GW signal strongly depends also on $\mathcal{R}$, the distance  between the gravitational source and the observer. This information is substantially cancelled out through the ratio in Eq. (\ref{eq:ERRREL}). The goal of the plot in Fig. \ref{fig:Fig2} aims at generally conveying the impact of the EC theory on the observations.

\begin{figure*}[t!]
    \centering
    \includegraphics[scale=0.62]{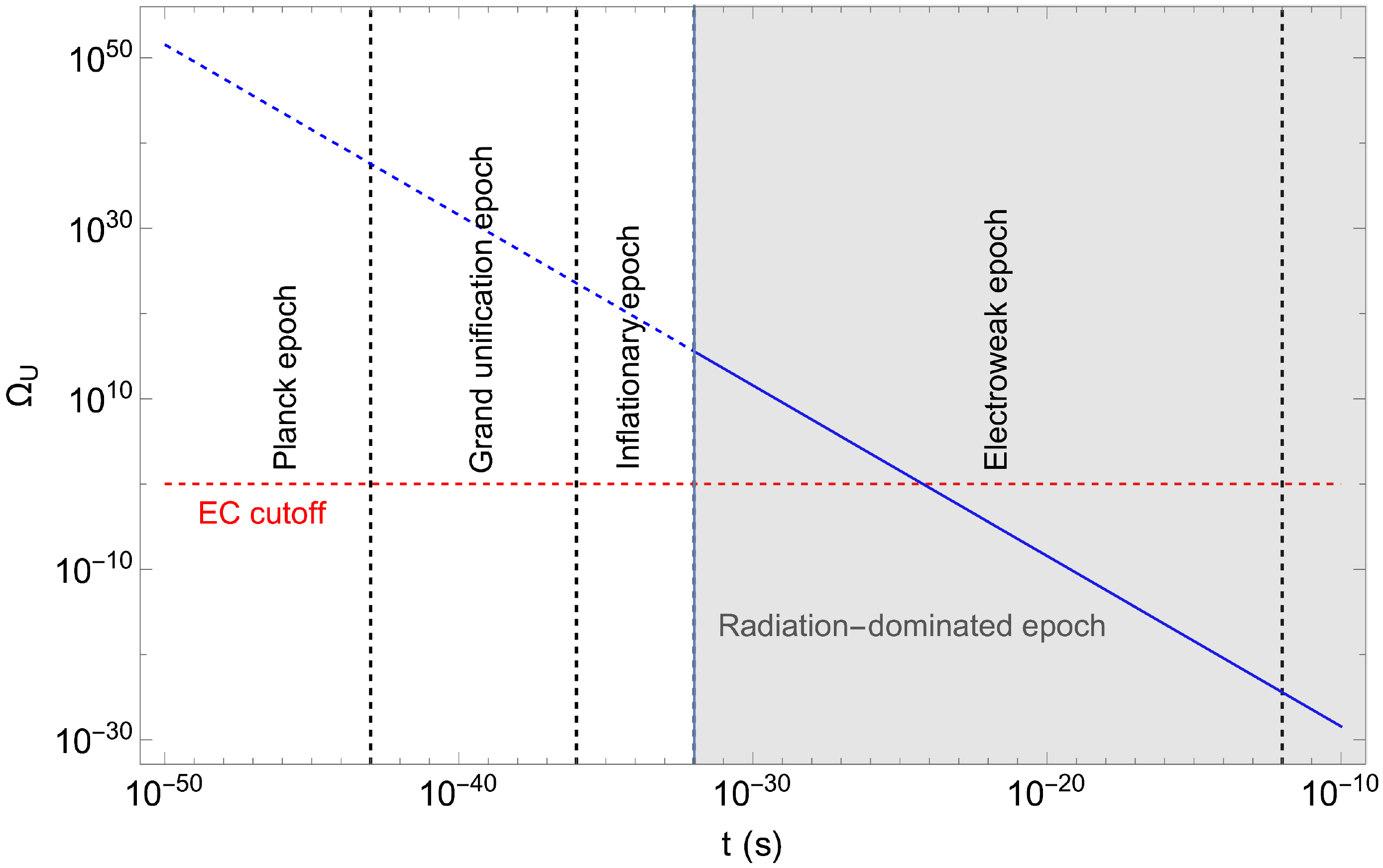}
    \caption{Plot of the density parameter $\Omega_U$, defined in  Eq. (\ref{eq:Omega-U}), as a function of the cosmic time $t$. The following cosmological eras have been considered:  \emph{Planck epoch} ($t<10^{-43}$ s), \emph{Grand unification epoch} ($10^{-43}\;{\rm s}<t<10^{-36}$ s), \emph{Inflation epoch}  ($10^{-36}\;{\rm s}<t<10^{-32}$ s), and \emph{Electroweak epoch} ($10^{-32}\;{\rm s}<t<10^{-12}$ s). The light grey-shaded area represents the \emph{radiation-dominated epoch} ($10^{-32}\;{\rm s} <t<1.5\times10^{12}$ s). The dashed red line is located at $\Omega_U=1$. The continuous blue line represents the trend of $\Omega_U$, whereas the dashed blue line refers to stages where quantum-gravity cosmological approaches should be invoked and hence Eq. (\ref{eq:FRW_density}) is not reliable. The blue line crosses the red dashed line at $t\simeq 6\times10^{-25}$ s.}
    \label{fig:Fig3}
\end{figure*}

Beside astrophysical compact objects, another fundamental testbed of the EC prediction power is represented by the GWs emitted from early stages of the formation of the Universe     \cite{Guzzetti2016,Bartolo2016,Caprini2018,Allahverdi2020}. Indeed, the  analysis of Eqs. (\ref{eq:Einstein-Cartan_equations}) and (\ref{Eq:S_alphabeta}) reveals that the physical regime in which EC theory predictions deviate significantly from GR expectations is ruled by the \emph{mass density factor} \cite{Hehl1976_fundations,Hehl1974,Arkuszewski1974}
\begin{equation} \label{eq:rho-EC}
\rho_{{\tiny EC}}= \dfrac{m_n^2 c^4}{8 \pi G \hbar^2} \simeq 1.21 \times 10^{57} \; {\rm kg}/{\rm m}^3,
\end{equation}
where $m_n=1.67\times10^{-27}$ kg is the mass of the neutron. Equation (\ref{eq:rho-EC}) represents a \emph{cutoff parameter for the EC model}, in the sense that for densities of the order of $\rho_{EC}$ the spin effects become the dominant source of the gravitational field. We can evaluate the epoch when the Universe has reached a density of the order of $\rho_{EC}$. If we consider the \emph{radiation-dominated era} (which starts when the cosmic time $t\simeq10^{-32}$ s  and extends until $t\simeq1.5\times 10^{12}$ s or equivalently $t\simeq4.7\times10^{4}$ yr \cite{Weinberg2008}) of the Friedmann-Robertson-Walker (FRW) cosmological framework, the density of the Universe can be described by \cite{Wald,MTW}
\begin{equation} \label{eq:FRW_density}
\rho(t)=\frac{3}{32\pi G t^2}.   
\end{equation}
In accordance with the cosmological approach, we can define the \emph{density parameter} 
\begin{equation} \label{eq:Omega-U}
    \Omega_U(t) \equiv \dfrac{\rho(t)}{\rho_{EC}},
\end{equation}
which for values greater than the unity indicates the importance of the EC model in the description of the Universe evolution. In Fig. \ref{fig:Fig3} we provide the plot of the function $\Omega_U(t)$ during the first moments after the Big Bang (assumed to be at $t=0$ s). Our estimation shows that in the first $10^{-32}$ seconds the quantum effects are very strong ($\Omega_U\gtrsim4\times10^{15}$) and play a fundamental role in determining the Universe dynamics. This scenario is well-known in the literature, and a cosmological quantum-gravity framework must be invoked in order to provide an accurate description \cite{Calcagni2013,Agullo2013,Alesci2013,Li2018,Calcagni2020}. From our calculations, we  infer that $\Omega_U=1$ at the cosmic time $t\simeq 6\times 10^{-25}$ s, suggesting that we can resort to the EC model at the beginning of the radiation-dominated epoch (i.e., for $10^{-30}\; {\rm s}\lesssim t \lesssim 10^{-23}$ s, which leads to  $10^{11}\lesssim \Omega_U \lesssim 10^{-2}$). The formalism developed in this paper could  be in principle exploited to extract fundamental information regarding the GWs from the early Universe in order not only to better understand its formation process, but also to tighter constraints on different cosmological models and gravity theories. However, its application to cosmological events demands that the following main improvements are taken into account: (1) the linearized theory should be superseded  since  EC equations must be expanded around a curve FRW-like background; (2) the stress-energy tensor  of the cosmological fluid filling the Universe should be considered.  

Our estimations, albeit very simple, suggest, together with theoretical developments, more and more the route of the future observations towards a deep understanding of gravity at quantum level.

\section*{Acknowledgements}
The authors thank the anonymous referee  for the useful comments. The authors are grateful to Professor L. Stella for fruitful discussions. E. B. is grateful to Professor Friedrich W. Hehl, Professor Vladimir N. Ponomarev, and Professor Yu N. Obukhov for useful explanations about some topics discussed in this paper. E. B. and V. D. F. thank the Silesian University in Opava and the International Space Science Institute in Bern for hospitality and support. E. B. and V. D. F. are grateful to Gruppo Nazionale di Fisica Matematica of Istituto Nazionale di Alta Matematica for support. V. D. F. acknowledges the support of INFN {\it sez. di Napoli}, {\it iniziative specifiche} TEONGRAV.

\bibliography{references}
\end{document}